\documentclass[11pt]{article}
\usepackage{amssymb,amsmath} 
\usepackage{graphicx}
\usepackage{hyperref}

\def\pl{\partial}

\def\={\equiv}

\newcommand{\xt}{(\3x,t)}

\newcommand{\zzz}{{\rm z}}

\newcommand{\ox}{(\3x)}

\newtheorem{thm}{Theorem}

\newcommand{\bib}{\bibitem}

\newcommand{\nt}{\notag}
\newcommand{\ci}{\cite}
\newcommand{\lab}{\label}

\newcommand{\eq}{\eqref}

\newcommand{\bx}[1]{\boxed{\ #1\ }}

\newcommand{\lp}{\left(}
\newcommand{\rp}{ \right)}
\newcommand{\lb}{ \left[}
\newcommand{\rb}{\right]}

\newcommand{\LB}{\left\lbrace}
\newcommand{\RB}{\right\rbrace}

\newcommand{\harr}[1]{\smash{\mathop{\hbox to .5in{\ \rightarrowfill\ }}
      \limits^{#1}}}

\newcommand{\0}[1]{{(#1)}}

\newcommand{\2}[1]{{\tilde #1}}
\newcommand{\3}[1]{{\boldsymbol #1}}

\newcommand{\bh}[1]{{\boldsymbol{\hat #1}}}

\newcommand{\6}[1]{_{\scriptscriptstyle#1}}

\newcommand{\8}{\infty}

\def\a{\alpha} 
\def\b{\beta}

\def\d{\delta} 
\def\e{\varepsilon} 

\def\f{\phi} 
 
\def\g{\gamma}
\def\h{\eta}

\def\l{\lambda}

\def\o{\omega} 
\def\p{\pi} 

\def\q{\theta} 
\def\vq{\vartheta} 
\def\r{\rho}
\def\vr{\varrho} 
\def\s{{\sigma}} 

\def\t{\tau} 
 
\def\x{\xi}
\def\y{\psi} 
\def\z{\zeta}

\def\F{\Phi}

\newcommand{\db}{{\,{\rm d}\kern-1.6ex-}}
\newcommand{\dir}{{\pl\kern-1.2ex {/}}}
\newcommand{\dd}{{\rm d}}

\newcommand{\app}{\approx} 
\newcommand{\cc}[1]{{{\mathbb C\hskip.5pt}^{#1}}}

\newcommand{\erf}{{\,\rm erf \,}}
\newcommand{\erfc}{{\,\rm erfc \,}}

\newcommand{\grad}{\nabla}

\newcommand{\ie}{{\it i.e., }}

\def\iff{\ \Leftrightarrow\ }
\newcommand{\im}{{\,\rm Im}\ }  
\newcommand{\imp}{\ \Rightarrow\ }
\newcommand{\inv}{^{-1}}

\newcommand{\ir}{\int_{-\infty}^\infty}  

\newcommand{\lra}{\leftrightarrow}

\newcommand{\plra}{\pl^{\kern-1.25ex^\lra}}
\newcommand{\qq}{\quad} 
\newcommand{\qqq}{\qquad} 
\newcommand{\re}{{\,\rm Re}\  }   

\newcommand{\rr}[1]{{{\mathbb R}^{#1}}}

\newcommand{\sgn}{{\,\rm Sgn \,}}
 
\newcommand{\sr}{\sqrt}

\newcommand{\supp}{{\rm supp \,}}
\newcommand{\sv}[1]{\vskip#1ex}

\def\XXint#1#2#3{{\setbox0=\hbox{$#1{#2#3}{\int}$}
     \vcenter{\hbox{$#2#3$}}\kern-.5\wd0}}

\def\bib#1{\bibitem[#1]{#1}}

\begin{document}

\title{Generalized Huygens principle with pulsed-beam wavelets}

\author{Thorkild Hansen, Seknion Inc., Boston, MA\\
Gerald Kaiser, Signals \& Waves,  Austin, TX
}

\maketitle

\begin{abstract} \noindent
Huygens' principle states that the solution of the wave equation radiated by a bounded source can be represented outside the source region as a superposition of spherical `Huygens wavelets' radiated by secondary point-sources on a surface enclosing the primary source. This was originally proposed as a \sl geometrical \rm explanation of wave propagation, but as such it is conceptually problematic because the spherical wavelets propagate equally in all directions, thus implying that the wave propagates \sl backwards \rm (toward the source) as well as forwards. We propose a solution to this problem by generalizing the idea of Huygens wavelets.  Choosing the surface to be the sphere $S_R$ of radius $R$, we show that the Huygens representation of the exterior wave can be continued analytically to a complex radius $\a=R+ia$. For any real unit vector $\bh n$, the complex vector $\a\bh n$ is shown to represent a \sl real \rm disk of radius $a$ tangent to $S_R$ at the point $R\bh n$.  The complex sphere $S_\a$ consisting of all such vectors $\a\bh n$ is therefore equivalent to a real \sl tangent disk bundle \rm with base $S_R$.  Just as the points $R\bh n\in S_R$ radiate spherical wavelets, so do the tangent disks $\a\bh n\in S_\a$ radiate well-focused  \sl pulsed-beam wavelets  \rm  propagating in the outward direction  $\bh n$. The analytically continued Huygens formula can be given the following real interpretation: the interior wave radiated by the source is \sl intercepted \rm by the set of tangent disks $\a\bh n$, which then re-radiate it as a set of outgoing pulsed beams. The original wave is thus represented in the exterior as a superposition of pulsed beams emanating from disks tangent to the sphere $S_R$, and the coefficients are interpreted as local reception amplitudes by the disks. 
The generalized principle is a \sl completeness relation \rm for pulsed-beam wavelets, enabling a \sl pulsed-beam representation \rm of radiation fields. Since the new wavelets can be \sl focused \rm by increasing the disk radius $a$, our construction solves the directionality problem of Huygens' original construction. Furthermore, it leads to substantial gains in the efficiency of computing radiation fields. Only pulsed beams propagating toward the observer need to be included and the rest can be ignored while incurring little error.  This leads to a significantly \sl compressed \rm representation of radiation fields, with the compression controlled by the disk radius $a$. We confirm these results by numerical simulations.
\end{abstract}

\tableofcontents

\section{Huygens principle for time-harmonic waves}\label{S:TimeHarm}

Consider a time-harmonic source $\vr$ of frequency $\o$ supported in a bounded volume $V\subset\rr3$:
\begin{align*}
\vr\xt=e^{-i\o t}\vr_\o\ox,\qq \supp\vr_\o\subset V.
\end{align*}
The wave radiated in free space and observed at the \sl reception event \rm $(\3x_r, t_r)$ is
\begin{align*}
F(\3x_r, t_r)=e^{-i\o t_r}F_\o(\3x_r)\ \ \hbox{with}\ \ 
F_\o(\3x_r)=\int\dd\3x\,G_\o(\3x_r-\3x)\vr_\o\ox,
\end{align*}
where 
\begin{align}\lab{Go}
G_\o(\3r)=\frac{e^{i\o r}}r,\qq r=|\3r|
\end{align}
is the outgoing fundamental solution of the Helmholtz equation:
\begin{align}\lab{fund0}
(\grad^2+\o^2) G_\o\ox=-4\p\d\ox.
\end{align}
We are using units in which the constant wave propagation speed $c=1$, so the wave number is $k\equiv\o/c=\o$.

Let $S$ be a smooth surface containing $V$ in its interior. Then Green's second identity, combined with \eq{fund0}, shows that $F_\o$ is given in the \sl exterior \rm of $S$ by
\begin{align}\lab{ext}
F_\o(\3x_r)=-\frac1{4\p}\int_S\dd S\ox\,G_\o(\3x_r-\3x)\plra_n F_\o\ox
\end{align}
where $\dd S$ is the area measure on $S$, $\pl_n$ is the outward normal derivative at $\3x\in S$, and we have introduced the notation
\begin{align*}
g\ox\plra_n f\ox=g\ox\pl_n f\ox-\pl_n g\ox\,f\ox.
\end{align*}
Equation \eq{ext} is a precise expression of  \sl Huygens' principle \rm as formulated by Kirchhoff \ci{BC87, BW99}. It states that in the exterior region,  $F_\o(\3x_r)$ can be represented as a superposition of the spherical waves $G_\o(\3x_r-\3x)$, called \sl Huygens wavelets, \rm together with their normal derivatives. Hence the points $\3x\in S$ act as secondary sources which collectively form a surface source \sl equivalent \rm to the original source $\vr_\o$ in the exterior region.\footnote{The equivalent surface source consists of a \sl single layer \rm $\{G_\o\pl_n f\}$ and a \sl double (dipole) layer \rm $\{-\pl_n G_\o \, f\}$.
}

Equation \eq{ext} can be expressed as a condition on the fundamental solution $G_\o$ by letting $\vr_\o$ be a point source 
\begin{align*}
\vr_\o\ox=\d(\3x-\3x_e)
\end{align*}
with $\3x_e$ in the interior of $S$. This gives $F_\o\ox=G_\o(\3x-\3x_e)$, hence  \eq{ext} becomes
\begin{align}\lab{rep}
G_\o(\3x_r-\3x_e)
=-\frac1{4\p}\int_S\dd S\ox\,G_\o(\3x_r-\3x)\plra_n G_\o(\3x-\3x_e).
\end{align}
We call \eq{rep} the \sl Huygens reproducing relation \rm for $G_\o$. To recover \eq{ext}, multiply by a general source density $\vr_\o(\3x_e)$ supported inside $S$ and integrate over $\3x_e$.

\begin{figure}[ht]
\begin{center}
\includegraphics[width=4 in]{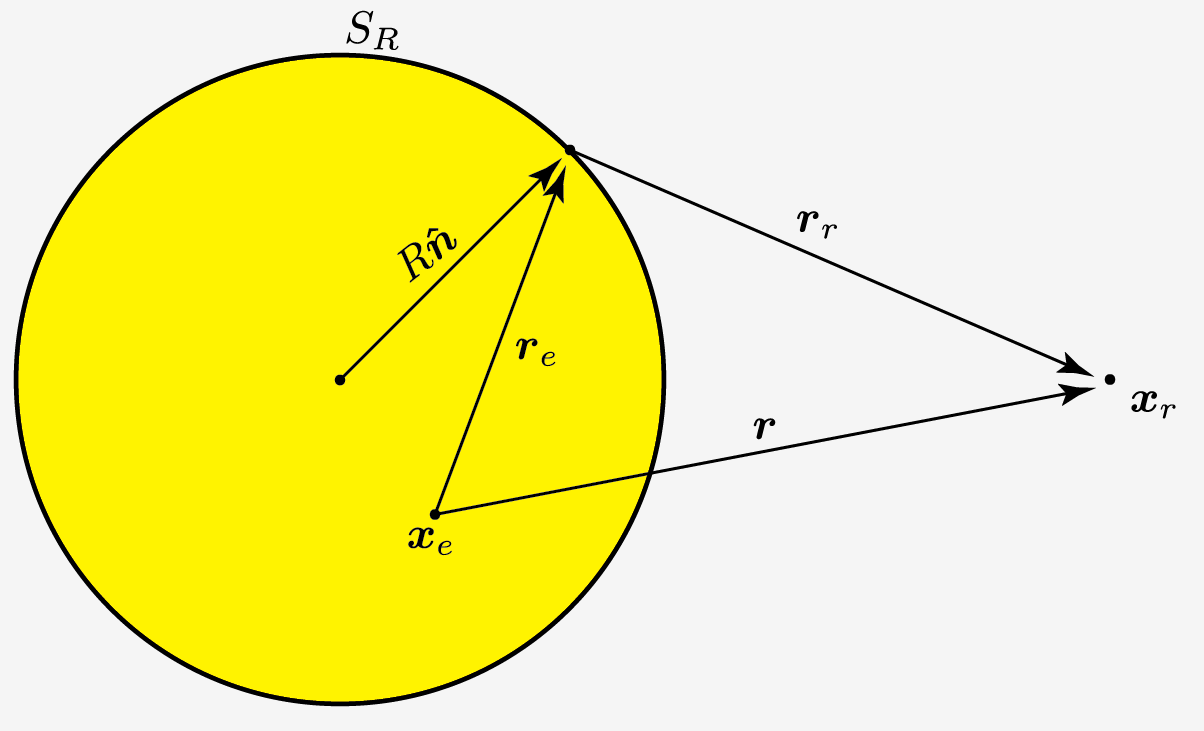}
\caption{\small The sphere $S_R$, the emission and reception points $\3x_e, \3x_r$, and the vectors $\3r_e, \3r_r$.}
\label{F:Fig_SR.png}
\end{center}
\end{figure}
We shall generalize Huygens' principle by continuing analytically in the integration variable $\3x$, and for this purpose it will be more convenient to work with \eq{rep} than \eq{ext}.  This will be done in the special case where $S$ is the sphere of radius $R$ centered at the origin,
\begin{align*}
S_R=\{\3x=R\bh n:\, \bh n\in S^2\},
\end{align*}
where $S^2$ denotes the unit sphere. Then
\begin{align*}
\dd S\ox=R^2\dd\bh n, \ \ \hbox{where}\ \ \dd\bh n=\sin\q\,\dd\q\,\dd\f
\end{align*}
is the area measure on $S^2$, hence \eq{rep} becomes
\begin{align}\lab{rep1}
G_\o(\3r)
=-\frac{R^2}{4\p}\int\dd\bh n\,G_\o(\3r_r)\plra\6R G_\o(\3r_e),\qq |\3x_e|<R<|\3x_r|
\end{align}
where
\begin{align}\lab{rer}
\3r_e=R\bh n-\3x_e,\qq \3r_r=\3x_r-R\bh n,\qq \3r=\3r_e+\3r_r=\3x_r-\3x_e
\end{align}
as seen in Figure \ref{F:Fig_SR.png}. The normal derivative $\pl_n=\bh n\cdot\grad$ has been replaced by the partial derivative $\pl\6R$, and
\begin{align}\lab{Gre}
G_\o(\3r_e)&=\frac{e^{i\o r_e}}{r_e}, \qq r_e=|\3r_e|,\qq \pl\6R r_e=\frac{R-\bh n\cdot\3x_e}{r_e}\\
G_\o(\3r_r)&=\frac{e^{i\o r_r}}{r_r}, \qq r_r=|\3r_r|,\qq \pl\6R r_r=\frac{R-\bh n\cdot\3x_r}{r_r}.
\nt
\end{align}
We shall complexify the points $R\bh n$ of $S_R$ by complexifying $R$ and proving that this gives an analytic continuation of the distances $r_e$ and $r_r$, hence of the right side in \eq{rep1}. In the next section we show that this procedure has a surprising and beautiful geometric interpretation in \sl real \rm space.

\section{The complex sphere as a tangent disk bundle}\label{S:CxSph}

Let $\a=R+ia\in\4C$ with $a>0$, and consider the complexifications of the vectors \eq{rer},
\begin{align}
\3z_e&=\a\bh n-\3x_e=\3r_e+ia\bh n,\qq
\3z_r=\3x_r-\a\bh n=\3r_r-ia\bh n, \lab{zer}
\end{align}
regarded as analytic functions of $\a$. To continue $G_\o(\3r_e)$ and $G_\o(\3r_r)$ in \eq{rep1} to $\cc3$, we must continue the distances $r_e\,, r_r$ analytically in $\a$.  We will first explain the continuation of $r_r$ in detail and then derive the corresponding expressions for $r_e$.

The \sl complex distance \rm  from  $\a\bh n$ to $\3x_r$ is defined by
\begin{align}\lab{zeta}
\z_r=\sr{w}\ \ \hbox{where}\ \ w=\3z_r\cdot\3z_r=r_r^2-a^2-2ia\3r_r\cdot\bh n.
\end{align}
$\z_r$ will be regarded \sl in parallel \rm as an analytic function of $\3z_r\in\cc3$ and as a complex function of $\3x_r\in\rr3$ with $\a\bh n\in\cc3$ fixed.  Any analytic function $f(\3x_r)$ depending only on $r_r$ can be continued analytically to some domain in $\cc3$ by substituting $r_r\to\z_r$, and we shall regard this as a \sl deformation \rm 
\begin{align}\lab{deform}
f(\3x_r)\to f_{\a\bh n}(\3x_r)\equiv f(\3x_r-\a\bh n).
\end{align}
Since $f_{\a\bh n}$ is analytic in $\a$, this deformation \sl preserves solutions of differential equations \rm such as \eq{fund0}.\footnote{We shall extend this idea to spacetime, where it applies, in particular, to solutions of the wave equation and Maxwell's equations.
}
The deformation breaks the spherical symmetry of $r_r$. Coupled with a similar deformation of other variables such as $r_e$, this will provide a powerful mathematical tool for generating nontrivial and \sl interesting \rm solutions from simple spherical ones. Furthermore, the \sl singularities \rm of deformed solutions give rise to their deformed \sl sources \rm \ci{K3}. 

\begin{figure}[ht]
\begin{center}
\includegraphics[width=3 in]{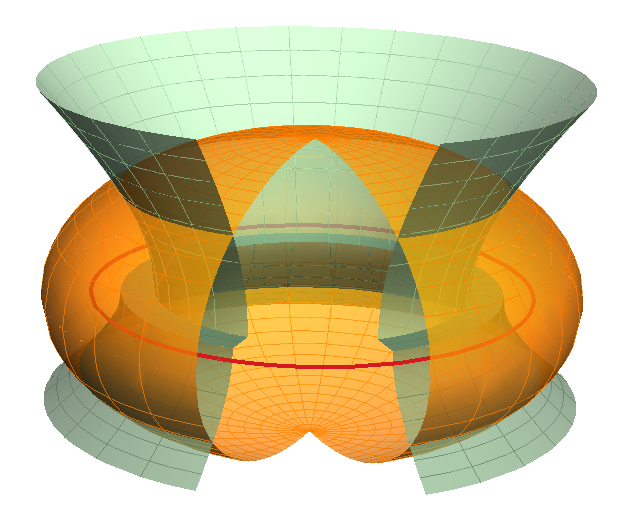}
\caption{\small The real and imaginary parts of $\z_r=\x_r-i\h_r$ form an oblate spheroidal coordinate system in $\rr3$ centered at $\3x_r=R\bh n$ with the $\zzz$-axis along $\bh n$. The third coordinate is $\f$, the standard azimuthal angle. The above plot shows  cut-away views of an oblate spheroid $\5O_{\x_r}$ with $\x_r=0.7a$, a semi-hyperboloid $\5H_{\h_r}$ with $\h_r=0.8a\, (\zzz>0)$ and another with $\h_r=-0.5a \,(\zzz<0)$. Also shown is the focal circle $\pl\5D(\a\bh n)$ with radius $a$, whose interior is the branch disk $\5D(\a\bh n)$.}
\label{F:Fig_OSCS_Color.png}
\end{center}
\end{figure}

Being defined in terms of the complex square root, $\z_r$ is double-valued. To make it single-valued, a branch cut must be introduced and a branch chosen. In the complex variable $w\in\4C$, we choose the standard branch cut of $\sr{w}$ along the negative real axis $w\le 0$. But
\begin{align*}
w\le 0&\iff \LB r_r\le a,\ \ \bh n\cdot\3r_r=0\RB\iff \LB|\3x_r-R\bh n|\le a,\ \ \bh n\cdot(\3x_r-R\bh n)=0\RB,
\end{align*}
hence the branch cut of $\z_r$ as a function of $\3x_r\in\rr3$ with $\a\bh n\in\cc3$ fixed is
\begin{align}\lab{Disk0}
\5D(\a\bh n)&=\{\3x_r: r_r\le a,\  \bh n\cdot\3r_r=0\}\\
&=\{\3x_r: |\3x_r-R\bh n|\le a,\ \bh n\cdot (\3x_r-R\bh n)=0\}.\nt
\end{align}
This is the disk of radius $a$ centered at $R\bh n$ and orthogonal to $\bh n$, \ie the disk of radius $a$ \sl tangent to the sphere $S_R$  at $R\bh n$. \rm As $a\to 0$, $\5D(\a\bh n)$ shrinks to the one-point set $\{R\bh n\}$ and $\z_r\to\pm r_r$. We choose the branch with
\begin{align*}
\re\z_r\ge 0,\ \ \hbox{so that}\ \ \z_r\to r_r \ \ \hbox{as}\ \  a\to0.
\end{align*}
Define the real and imaginary parts of $\z_r$ by
\begin{align}\lab{xh}
\bx{\z_r=\x_r-i\h_r}
\end{align}
so that with our choice of branch, 
\begin{align*}
\x_r\ge 0\ \ \hbox{and}\ \ \sgn\h_r=\sgn(\3r_r\cdot\bh n)
\end{align*}
by \eq{zeta}. Since $w\le 0$ on $\5D(\a\bh n)$, $\z_r$ is imaginary there; hence the branch cut can be characterized as
\begin{align}\lab{x0}
\5D(\a\bh n)=\{\3x_r: \x_r=0\}.
\end{align}
Choosing cylindrical coordinates $(\r,\f,\zzz)$ with the origin at $R\bh n$ and the $\zzz$-axis along $\bh n$, \eq{zeta} and  \eq{xh} give
\begin{align*}
r_r^2-a^2=\x_r^2-\h_r^2\ \ \hbox{and}\ \ a\bh n\cdot\3r_r=a\zzz=\x_r\h_r
\end{align*}
hence
\begin{align*}
a^2\r^2&=a^2r_r^2-a^2\zzz^2=a^2(a^2+\x_r^2-\h_r^2)-\x_r^2\h_r^2=(a^2+\x_r^2)(a^2-\h_r^2).
\end{align*} 
Thus $(\x_r,\h_r)$ are related to the cylindrical coordinates  $(\r,\zzz)$ by
\begin{align}\lab{xhrz}
\bx{a\r=\sr{a^2+\x_r^2}\sr{a^2-\h_r^2},\qq \ a\zzz=\x_r\h_r.}
\end{align}
This implies the following important inequalities: 
\begin{align}\lab{ineq}
-a\le\h_r\le a\ \ \hbox{and}\ \  0\le\x_r\le r_r
\end{align}
where the second one follows from $\x_r^2=r_r^2-(a^2-\h_r^2)$. Also by \eq{xhrz},
\begin{align*}
\frac{\r^2}{a^2+\x_r^2}+\frac{\zzz^2}{\x_r^2}=1\ \ \hbox{and}\ \ 
\frac{\r^2}{a^2-\h_r^2}-\frac{\zzz^2}{\h_r^2}=1.
\end{align*}
This proves that the level surfaces of $\x_r$ and $\h_r$ are
\begin{align}\lab{OH}
\5O_{\x_r}&=\LB\3x_r: \frac{\r^2}{a^2+\x_r^2}+\frac{\zzz^2}{\x_r^2}=1\RB,\qq \x_r>0\\
\5H_{\h_r}&=\LB\3x_r: \frac{\r^2}{a^2-\h_r^2}-\frac{\zzz^2}{\h_r^2}=1,
\ \zzz\h_r\ge 0\RB,\qq 0<\h_r^2<a^2.\nt
\end{align}
The level surfaces of $\x_r$ are the \sl oblate spheroids \rm $\5O_{\x_r}$ and those of $\h_r$ are the \sl semi-hyperboloids \rm $\5H_{\h_r}$.  The restriction $ \zzz\h_r\ge 0$ follows from $a\zzz=\x_r\h_r$ and $\x_r>0$. As $\x_r\to 0$, $\5O_{\x_r}$ shrinks to the branch disk $\5D(\a\bh n)$  \eq{x0}. It can be shown \ci{K3} that the families $\5O_{\x_r}$ and $\5H_{\h_r}$ are mutually orthogonal, forming an \sl oblate spheroidal coordinate system  \rm deforming the spherical coordinates $(r_r, \q_r,\f_r)$. They all share a common \sl focal circle,\rm\footnote{Its physical significance is that it consists entirely of focal points of both $\5O_{\x_r}$ and $\5H_{\h_r}$.
}
which is the boundary of the branch disk:
\begin{align}\lab{Cia}
\pl\5D(\a\bh n)=\{\3x_r: r_r=a,\, \bh n\cdot\3r_r=0\}=\{\3x_r: \z_r=0\}.
\end{align}
The last equality shows that $\pl\5D(\a\bh n)$ is the set of all \sl branch points \rm of $\z_r$. 
Whereas $f\0w=\sr{w}$ has a branch \sl point \rm at $w=0$, $\z_r(\3x_r)=\sr{(\3x_r-\a\bh n)^2}$ has a branch \sl  circle. \rm Figure \ref{F:Fig_OSCS_Color.png} shows $\pl\5D(\a\bh n)$ and examples of $\5O_{\x_r}$ and $\5H_{\h_r}$.

\begin{figure}[ht]
\begin{center}
\includegraphics[width=4 in]{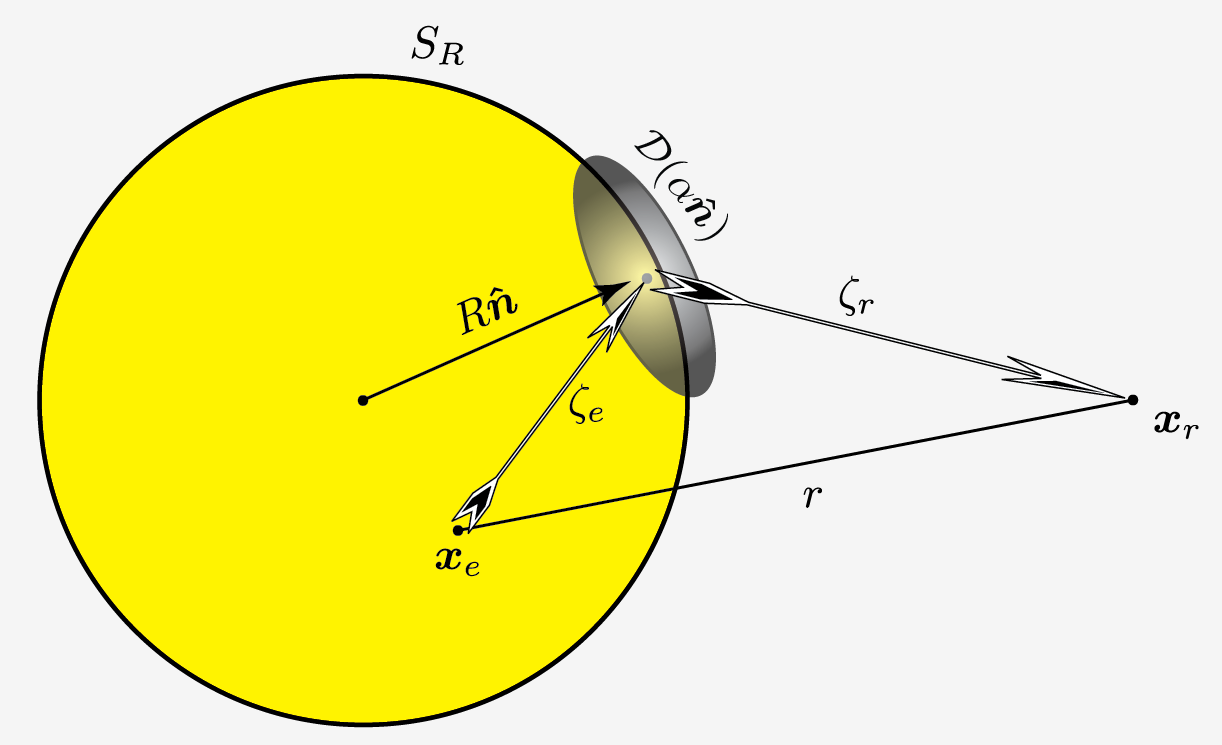}
\caption{\small The complex point $\a\bh n$ represents the real disk $\5D(\a\bh n)$ tangent to the sphere $S_R$ at $R\bh n$. The complex distances $\z_e$ (from $\3x_e$ to $\5D(\a\bh n)$) and
$\z_r$ (from $\5D(\a\bh n)$ to $\3x_r$) are depicted schematically, emphasizing their \sl directed \rm nature as explained in the text.}
\label{F:Fig_Sa.png}
\end{center}
\end{figure}

Since $(-\3z_r)^2=\3z_r^2 $, $\z_r$ is even as a function of $\3z_r\in\cc3$. 
However, it is \sl not \rm even as a function of $\3r_r$ alone. Instead, we have
\begin{align*}
\z_r(-\3r_r-ia\bh n)=\z_r(\3r_r+ia\bh n)=\z_r(\3r_r-ia\bh n)^*.
\end{align*}
The last relation is a \sl reality condition \rm or \sl Hermiticity property \rm on the complex function $\z_r(\3z_r)$, and it requires our choice of branch cut $\x_r\ge 0$:
\begin{align}\lab{reality}
\z_r(\3z_r^*)=\z_r(\3z_r)^*.
\end{align}
We now use this to define the analytic continuation of $r_e$ by
\begin{align*}
\z_e=\sr{(\3r_e+ia\bh n)^2}=\lp\sr{(\3r_e-ia\bh n)^2}\rp^*
=(\x_e-i\h_e)^*=\x_e+i\h_e.
\end{align*}
This shows that $\z_e$ and $\z_r$ are \sl directed \rm distances. Their sign difference indicates that $\a\bh n$ is a \sl receiver \rm for the wave propagating from $\3x_e$ and an \sl emitter \rm for the wave propagating to $\3x_r$, as illustrated in Figure \ref{F:Fig_Sa.png}. The sign difference is significant because $\pm\5D(\a\bh n)$ have \sl opposite orientations. \rm 

\begin{figure}[ht]
\begin{center}
\includegraphics[width=3 in]{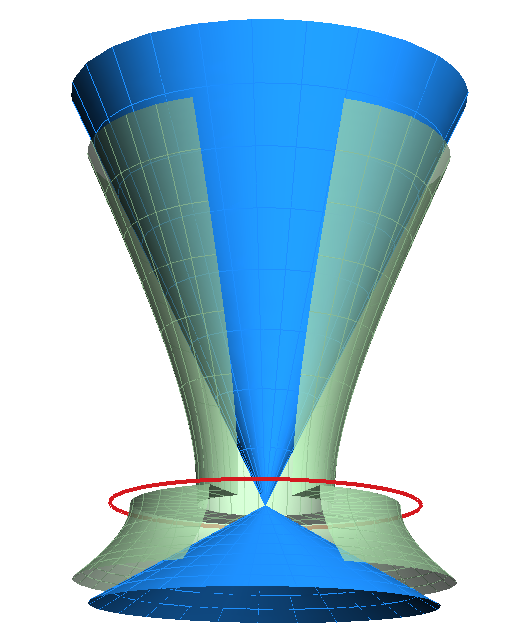}
\caption{\small The semi-hyperboloids $\5H_{\h_r}$ (above) and $\5H_{\h_e}$ (below) for $\h_r=0.9 a$ and $\h_e=0.5a$. Also shown are the asymptotic cones $\5C_{\vq_r}, \5C_{\vq_e}$ and the focal circle $\pl\5D(\a\bh n)$. The cones make angles $\vq_r$ and $\vq_e$ with $\bh n$ and $-\bh n$, respectively, given by \eq{qre}. The Huygens relation in the time domain will favor pulsed beams with $\h_r>\h_e$, which means that the wave is \sl focused \rm into narrower propagation hyperboloids $\5H_{\h_r}$ (asymptotic to the diffraction cones $\5C_{\vq_r}$) upon being received along $\5H_{\h_e}$ and re-emitted by $\5D(\a\bh n)$.}
\label{F:Fig_Cones2_Color.png}
\end{center}
\end{figure}

As functions of $\3x_e$ for fixed $\a\bh n\in\cc3$, $\x_e$ and $\h_e$ have the same properties as $\x_r$ and $\h_r$ except that the $\zzz$-axis is now along $-\bh n$ due to the opposite orientation of $\5D(\a\bh n)$. For example, the level surfaces of $\x_e$ are oblate spheroids and those of $\h_e$ are semi-hyperboloids with 
\begin{align*}
\x_e\ge 0,\qq \sgn\h_e=\sgn(\bh n\cdot\3r_e).
\end{align*}
However, it is clear from Figure \ref{F:Fig_SR.png} that while $\h_r$ can have any value in $[-a,a]$, every emission point $\3x_e$ in the interior of $S_R$ must have
\begin{align*}
\bh n\cdot\3r_e>0, \ \ \hbox{hence}\ \ 0<\h_e\le a.
\end{align*}
It can be shown that the exact bounds on $\h_e$ as $\bh n$ varies over $S^2$ are
\begin{align}\lab{etaBounds}
\g a\le\h_e\le a,\ \ \hbox{where}\ \ \g=\sr{1-\frac{|\3x_e|^2}{|\a|^2}}. 
\end{align}
It is clear that $\g$ can depend only on $|\3x_e|$ since the minimum of $\h_e$ must be spherically symmetric. In particular, 
\begin{align}\lab{gamma0}
0\le |\3x_e|<R\imp 0<\g\le 1,\qqq  \g\60\equiv \lim_{|\3x_e|\to R}\g=\frac a{\sr{R^2+a^2}}<1,
\end{align}
and
\begin{align*}
\3x_e=\30\imp \h_e=a\ \ \hbox{for all}\ \ \bh n,
\end{align*}
which is obvious since
\begin{align*}
\3x_e=\30\imp\3z_e=\a\bh n\imp\z_e=\a\imp\x_e=R,\ \h_e=a.
\end{align*}
The functions $\h_r$ and $\h_e$ will play an important role, and it is helpful to interpret them geometrically. From \eq{OH} it follows that $\5H_{\h_r}$ is asymptotic to the \sl cone \rm $\5C_{\vq_r}$ making an angle $\vq_r$ with the positive $\zzz$-axis and $\5H_{\h_e}$ is asymptotic to the cone $\5C_{\vq_e}$ making an angle $\vq_e$ with the negative $\zzz$-axis, where
\begin{align}\lab{qre}
a\cos\vq_r=\h_r\ \ \hbox{and}\ \ a\cos\vq_e=\h_e\,.
\end{align}
See Figure \ref{F:Fig_Cones2_Color.png}.
Hence \eq{etaBounds} can be restated as
\begin{align}\lab{thetaBounds}
\bx{0\le\vq_e\le\sin\inv\frac{|\3x_e|}{|\a|}}
\end{align}
while  $0\le\vq_r\le\p$.

The parameters $\vq_r, \vq_e$ are deformations of the spherical coordinates $\q_r, \q_e$ of $\3r_r, \3r_e$. A similar interpretation exists for $\x_r$ and $\x_e$ as deformations of the radial coordinates $r_r, r_e$: the oblate spheroid $\5O_{\x_r}$ containing $\3x_r$ is tangent to the sphere $S_{\x_r}$ of radius $\x_r$ at the north and south poles, and the same goes for $\5O_{\x_e}$ and $S_{\x_e}$; that explains why $\x_r\le r_r$ and $\x_e\le r_e$. These observations provide a complete \sl real \rm geometric interpretation of the complex distances $\z_r$ and $\z_e$ in $\rr3$. As an intuitive aid to understanding the idea, think of $\z_r$ as the `distance' between the disk $\5D(\a\bh n)$ and the point $\3x_r$. Its complex nature reflects the fact that no single \sl real \rm number can characterize this distance, and that the distances from $\3x_r$ to points on $\5D(\a\bh n)$ depend on the inclination of the disk, which can be parameterized by $\vq_r$ or $\h_r$. Hence $\z_r$ is not spherically symmetric, like $r_r$,  but cylindrically symmetric around $\bh n$.

The functions $\x_r, \h_r$ simplify if the observer is far from the disk:
\begin{align}\lab{far0}
r_r\gg a&\imp \z_r=\sr{r_r^2-a^2-2ia\3r_r\cdot\bh n}\sim r_r-ia\bh r_r\cdot\bh n\\
&\imp\x_r\sim r_r,\qq \h_r\sim a\cos\q_r\ \ \hbox{where}\ \ \cos\q_r\equiv \bh r_r\cdot\bh n. \nt
\end{align}
In particular, note that $\vq_r\sim\q_r$ as expected.  Hence the spheroids $\5O_{\x_r}$  can be approximated by the spheres $S_{r_r}$ and the semi-hyperboloids $\5H_{\h_r}$ by their asymptotic cones $\5C_{\vq_r}$. The deformed variables $(\x_r,\vq_r)$ are thus restored to their original values $(r_r,\q_r)$. On the other hand, if the observer is far from the \sl sphere, \rm then
\begin{align*}
|\3x_r|\gg R\imp r_r=|\3x_r-R\bh n|\sim |\3x_r|-R\bh x_r\cdot\bh n.
\end{align*}
The \sl far-zone approximation \rm assumes that the observer is far from both the disk and the sphere, which can be stated succinctly as follows:
\begin{align*}
|\3x_r|\gg |\a|\imp \z_r=\sr{|\3x_r|^2+\a^2-2\a\3x_r\cdot\bh n}\sim |\3x_r|-\a\bh x_r\cdot\bh n
\end{align*}
or
\begin{align}\lab{far}
|\3x_r|\gg |\a|\imp \x_r\sim r_r\sim |\3x_r|-R\cos\q_r\ \ \hbox{and}\ \ \h_r\sim a\cos\q_r.
\end{align}

In the engineering literature \ci{N86}, the set 
\begin{align}\lab{Sa}
S_\a=\{\a\bh n\in\cc3: \bh n\in S^2\}
\end{align}
is called the \sl complex sphere\rm\,\footnote{The term would be more appropriately applied to
\begin{align}\lab{Sa1}
\2S_\a=\{\3z\in\cc3: \3z\cdot\3z=\a^2\}.
\end{align}
Since $S_\a$ has real dimension 2 for $\a\ne 0$ while $\2S_\a$ has real dimension 4 (complex dimension 2), $S_\a$ is a proper subset of $\2S_\a$. 
}
of radius $\a$ in $\cc3$. The correspondence 
\begin{align}\lab{corr}
\a\bh n\in\cc3\lra \5D(\a\bh n)\subset\rr3
\end{align}
establishes a complete equivalence between complex \sl points \rm and real \sl disks \rm  (where a `disk' with radius $a=0$ is by definition a point). Under this equivalence, $S_\a$ corresponds to the set of all disks of radius $a$ tangent to $S_R$, which is a \sl tangent disk bundle \rm with base $S_R$:
\begin{align}\lab{Ta}
T_a(S_R)=\{\5D(\a\bh n):\, \bh n\in S^2\},\qq \a=R+ia.
\end{align}

\section{Generalized principle for time-harmonic waves}\label{S:AnHuyTimeHarm}

We can now continue  \eq{rep} to complex space by extending \eq{Gre} to
\begin{align}\lab{Go1}
\2G_\o(\3z_e)&=\frac{e^{i\o\z_e}}{\z_e},\qq \3z_e=\a\bh n-\3x_e,\qq \z_e=\sr{\3z_e\cdot\3z_e}\\
\2G_\o(\3z_r)&=\frac{e^{i\o\z_r}}{\z_r},\qq \3z_r=\3x_r-\a\bh n,\qq \z_r=\sr{\3z_r\cdot\3z_r}.\nt
\end{align}
If the observer is far from the disk, \eq{far0} gives 
\begin{align}\lab{Go2}
r_r\gg a\imp \2G_\o(\3z_r)\sim\frac{e^{i\o r_r}}{r_r}\,e^{\o a\cos\q_r},
\end{align}
where we have used $\z_r\sim r_r-ia\cos\q_r\sim r_r$ in the denominator.
Thus $\2G_\o$, viewed as a function of $\3r_r\in\rr3$, has a \sl radiation pattern \rm \ci{HY99}
\begin{align*}
\5F_\o(\q_r)=e^{\o a\cos\q_r}.
\end{align*}
For $\o>0$, this is the pattern of a \sl beam propagating in the direction of $\bh n$, \rm while for $\o<0$ the beam propagates in the direction of $-\bh n$. The larger $\o a$,\footnote{Recall that $c=1$, so $\o a=ka=2\p a/\l$ where $k$ is the wave number and $\l$ is the wavelength. Thus  $\o a$ can be interpreted as the \sl number of wavelengths in the circumference of $\pl\5D(\a\bh n)$. \rm
}
the sharper the beam. Note further that these beams are very special in that they have \sl  no sidelobes. \rm That makes them especially useful in applications such as communications and remote sensing. Analyticity in $\3z_r$ combines with \eq{fund0} to give
\begin{align}\lab{supp0}
(\grad_r^2+\o^2)\2G_\o(\3x_r-\a\bh n)=0\ \ \hbox{when}\ \ \3x_r\notin\5D(\a\bh n), 
\end{align}
where $\grad_r^2$ is the Laplacian with respect to $\3x_r$. This proves that the disk $\5D(\a\bh n)$ is the  \sl source \rm of the beam. Just as the Huygens wavelet $G_\o(\3x_r-R\bh n)$ is radiated by a point source $\d(\3x_r-R\bh n)$ at $\3x_r=R\bh n$, as seen from  \eq{fund0}, so is the beam $\2G_\o(\3x_r-\a\bh n)$ radiated by the branch disk $\5D(\a\bh n)$. This will be made more precise later, in the time domain. In the limit $a\to 0$, $\2G_\o(\3x_r-\a\bh n)$ becomes the spherical wavelet $G_\o(\3x_r-R\bh n)$.

This method of deforming spherical time-harmonic waves to beams was first introduced by Deschamps \ci{D71} and has become very popular in the engineering literature under the name \sl complex-source beams, \rm \ie beams due \sl formally \rm to a `point source' in $\cc3$, in our case $\a\bh n$, but interpreted physically as a real disk  \ci{KS71,F76,C81,F82}. Solutions to scattering problems where the incident field is a complex-source beam are readily obtained by analytically continuing solutions with a spherical incident field  \ci{CH89}. Complex-point \sl receivers \rm were first introduced in \ci{ZSB96} to model directed electroacoustic transducers in ultrasonics, and they have subsequently proven useful in cylindrical and spherical near-field scanning for both acoustic and electromagnetic fields \ci{H6,H9,H9A}.

An earlier application of complex distance was made in General Relativity by  Ted Newman and his collaborators \ci{NJ65,N65}, who used it to give simple derivations of spinning black holes with and without charge (Kerr and Kerr-Newman solutions) by deforming  known spherically symmetric solutions through analytic continuation.\footnote{The first derivation of a cylindrically symmetric solution of Einstein's equation was given by Roy Kerr in 1963 \ci{K63}. It was very complicated, which explains why it had taken 48 years to generalize Karl Schwarzschild's spherical solution. Newman's derivation, based on the complex distance, was a model of simplicity.
}

However, none of the above works actually \sl compute \rm the source of $\2G_\o$. This is not trivial because the singularities of $\2G_\o$ on $\5D(\a\bh n)$ are complicated by the branch cut: $\2G_\o$ is infinite on the focal circle $\5F(\a\bh n)$, where $\z_r=0$, and discontinuous on its interior. In \ci{K3}, the source $\2\d_\o$ of $\2G_\o$ is \sl defined \rm  by extending \eq{fund0} and \eq{supp0} to
\begin{align}\lab{2dz}
4\p\2\d_\o(\3x_r-\a\bh n)\equiv -(\grad_r^2+\o^2)\2G_\o(\3x_r-\a\bh n)
\end{align}
where $\grad_r^2$ is the \sl distributional \rm Laplacian with respect to $\3x_r$.
It is proved that $\2\d_\o$ is a generalized function supported on $\3x_r\in\5D(\a\bh n)$. In \ci{K4a}, it is shown that the analytically continued Coulomb potential
\begin{align*}
\2\F=\frac1\z,\qq \3z=\3x-i\3a\in\cc3,\qq \z=\sr{\3z\cdot\3z}
\end{align*}
generates a \sl real \rm electromagnetic field $(\3E,\3H)$ in the complex-analytic form
\begin{align*}
-\grad\2\F=\3E+i\3H.
\end{align*} 
which in turn identifies its source $\5D(\3a)$ as a spinning charged disk whose boundary moves at the speed of light. This is the flat-space version of the Kerr-Newman black hole, studied from a different viewpoint by Newman in \ci{N73}. This analysis is generalized to higher dimensions in \ci{K0}, where a rigorous connection between solutions of Laplace's equation in $\4R^{n+1}$ and the wave equation in $\4R^{n,1}$ (Minkowski space with $n$ space dimensions plus time) is established, generalizing earlier work by Garabedian \ci{G64}.

We are now ready to state and prove the analytic Huygens principle for time-harmonic waves.

\thm For given emission and reception points $\3x_e, \3x_r$ with $|\3x_e|<|\3x_r|$, the Huygens reproducing relation \eq{rep1} extends analytically to complex $R$ in the open set
\begin{align}\lab{A}
 A=\{\a\in\4C: \re\a>|\3x_e|,\ |\a|<|\3x_r|\},\qq \a=R+ia.
\end{align}
For $\a\in A$, it states that
\begin{align}\lab{rep2}
G_\o(\3x_r-\3x_e) =-\frac{\a^2}{4\p}\int\dd\bh n\,\2G_\o(\3x_r-\a\bh n)\plra\6\a \2G_\o(\a\bh n-\3x_e)
\end{align}
or
\begin{align}\lab{rep3}
\frac{e^{i\o r}}r
=-\frac{\a^2}{4\p}\,\int\dd\bh n\,\frac{e^{i\o\z_r}}{\z_r}\,\plra_\a \frac{e^{i\o\z_e}}{\z_e},\qq r=|\3x_r-\3x_e|.
\end{align}
\rm

\bf Proof: \rm Write \eq{rep2} as
\begin{align*}
\5L=\5R\0\a,
\end{align*}
where the left side $\5L$ is independent of $\a$ as noted. This reduces to \eq{rep1} for $\a=R$ with $|\3x_e|< R<|\3x_r|$. The right side $\5R\0\a$ is analytic as long as neither $\3x_e$ nor $\3x_r$ belong to any of the branch disks $\5D(\a\bh n)$. But the union of all these branch disks is the spherical shell
\begin{align}\lab{shell}
S_R^a=\bigcup_{\bh n\in S^2}\5D(\a\bh n)=\{\3x\in\rr3:\ R\le |\3x|\le\sr{R^2+a^2}=|\a|\},
\end{align}
hence $\3x_e$ must be in the interior of $S_R^a$ and $\3x_r$ in its exterior. This means that $\5R\0\a$ is analytic in $A$, and since it is constant on the line segment $A\cap\4R$, it must be constant throughout $A$.   $\blacksquare$

Equation \eq{rep2} can be interpreted physically as follows: $\2G_\o(\a\bh n-\3x_e)$ is the \sl reception amplitude \rm by the disk $\5D(\a\bh n)$ of the wave emitted by $\3x_e$, which in turn stimulates the \sl emission \rm of the complex-source beam $\2G_\o(\3x_r-\a\bh n)$ propagating to $\3x_r$. The spherical wave $G_\o(\3x_r-\3x_e)$ from $\3x_e$ to $\3x_r$ is thus represented as a sum of beams.

Equation \eq{rep3} can be further simplified by letting
\begin{align}\lab{zab}
\z_e\0\a=\sr{(\a\bh n-\3x_e)^2}\ \ \hbox{and}\ \ \z_r(\a')=\sr{(\3x_r-\a'\bh n)^2}
\end{align}
with $\a$ and $\a'$ independent. Then 
\begin{align}\lab{rep4}
\bx{\frac{e^{i\o r}}r
=\frac{\a^2}{4\p}\,\pl_{\a'\a}\int\frac{\dd\bh n}{\z_r\z_e}\,e^{i\o(\z_r+\z_e)}}
\end{align}
where
\begin{align*}
\pl_{\a'\a}=\{\pl_{\a'}-\pl_\a\}\bigm|_{\a'=\a}\!.
\end{align*}
Applying the derivatives gives a version of \eq{rep4} more suitable for numerical computations:
\begin{align}\lab{rep40}
\frac{e^{i\o r}}r
=\frac{\a^2}{4\p}\,\int\frac{\dd\bh n}{\z_r\z_e}\lb i\o(\z_r'-\z_e')-\frac{\z_r'}{\z_r}+\frac{\z_e'}{\z_e}  \rb e^{i\o(\z_r+\z_e)}
\end{align}
where
\begin{align}\lab{zrep}
\z_e'\equiv \pl_\a\z_e=\frac{\a-\bh n\cdot\3x_e}{\z_e},\qq \z_r'\equiv \pl_{\a'}\z_r=\frac{\a'-\bh n\cdot\3x_r}{\z_r}.
\end{align}

Let us note a symmetry of \eq{rep4}. Since the left side satisfies the Fourier transform reality condition $\1f(-\o)^*=\1f\0\o$, so must the right side; thus 
\begin{align}\lab{rep6}
\frac{e^{i\o r}}r=
\frac{{\a^*}^2}{4\p}\,\pl_{\a'^*\a^*}\int\frac{\dd\bh n}{\z_r^*\z_e^*}\,e^{i\o(\z_r^*+\z_e^*)}.
\end{align}
The branches defined by $\re\z_e\ge 0$ and $\re\z_r\ge 0$ satisfy the reality conditions \eq{reality}
\begin{align*}
\z_e\0\a^*=\z_e(\a^*)\ \ \hbox{and}\ \ \z_r(\a')^*=\z_r(\a'^*),
\end{align*}
hence the right side of \eq{rep6} is simply \eq{rep4} with $\a\to\a^*$ and $\a'\to\a'^*$. Since the set $A$ \eq{A} is symmetric under complex conjugation, this explains why \eq{rep6} and \eq{rep4} are consistent. That is, the right side of \eq{rep4} satisfies the \sl extended \rm reality condition
\begin{align}\lab{erc}
\1f(-\o, \a^*)^*=\1f(\o,\a).
\end{align}
Equation \eq{rep2} implies that the field radiated by an arbitrary source $\vr_\o(\3x_e)$ supported in $S_R$ is
\begin{align}\lab{Foxr}
F_\o(\3x_r) =\frac{\a^2}{4\p}\pl_{\a'\a}\int\dd\bh n\,\2G_\o(\3x_r-\a'\bh n)\2F_\o(\a\bh n)
\end{align}
where
\begin{align}\lab{Fan}
\2F_\o(\a\bh n)=\int\dd\3x_e\,\2G_\o(\a\bh n-\3x_e)\vr_\o(\3x_e)
\end{align}
is the analytic continuation of the radiated field $F_\o(R\bh n)$ from $S_R$ to $S_\a$. $\2F_\o(\a\bh n)$ can be interpreted as the \sl reception amplitude \rm of the radiation field by the disk $\5D(\a\bh n)$ \ci{ZSB96}. See also Section  \ref{S:RecAmp}, where this is proved in the time domain using a rigorous definition of pulsed-beam sources. Thus \eq{Foxr} has a simple physical interpretation: the field radiated by $\vr_\o$ is \sl intercepted by $\5D(\a\bh n)$ and re-radiated by  $\5D(\a'\bh n)$ \rm to give an identical field in the exterior, showing that $\vr_\o$ can be replaced by an equivalent source on the tangent disk bundle $T_a(S_R)$ given in \eq{Ta}.

Equation \eq{Foxr} gives the field radiated by $\vr_\o$ as a superposition of the complex-source beams $\2G_\o(\3x_r-\a\bh n)$ with source points $\a\bh n\in S_\a$.  The first exact representation of this type was obtained by Norris \ci{N86}, who expressed the field of a single real point source at the \sl origin \rm in terms of complex-source beams emanating from a sphere centered at the origin. Heyman \ci{H89} translated Norris' result into the time domain using the analytic-signal (positive-frequency) Fourier transform. Norris and Hansen subsequently generalized the result to arbitrary bounded sources, both in the frequency domain \ci{NH97} and time domain \ci{HN97}. 

However, the representations \ci{NH97,HN97} are very different from \eq{Foxr}. They express the \sl weights \rm of the complex-source beams in terms of the spherical-harmonic expansion coefficients of $\vr_\o$, which requires only the field and not its normal derivative.
On the other hand, since each of these coefficients involves an integration of the field over the entire sphere, it is not possible to express the weight of the complex-source beam emanating from $\a\bh n$ in terms of the incident field at that point, as in \eq{Foxr}. Hence the expansions in \ci{NH97} and \ci{HN97} are \sl nonlocal, \rm and consequently they do not have a straightforward physical interpretation like the one above.

An electromagnetic analog of \eq{Foxr} has been published in \ci{TPB7} and used in \ci{TPB7A} to accelerate the method of moments. 

The representations \eq{rep2} and \eq{Foxr} can be further generalized to surfaces $S$ other than spheres. It need not even be assumed that the source disks represented by the points of the analytically continued surface $\2S$ must be tangent to $S$. However, this more general analytic continuation is more difficult than extending a single parameter like $R$. It does not work for all `regular' surfaces\footnote{A \sl regular \rm surface is defined by Kellogg \ci{K67}; see also \ci[Chapter 2]{HY99}.
}
for which a real Huygens representation holds because the integral expression is not necessarily analytic in a sufficiently large domain. To obtain an analytic continuation for a surface $S$, it is necessary to ensure that (a) the integration avoids all branch cuts, and (b) the area measure of $S$, which involves a Jacobian, can be continued analytically. These topics will be considered in future work.

\section{Gaussian pulsed beams as Huygens wavelets}\label{S:HuyWavGPB}

\begin{figure}[ht]
\begin{center}
\includegraphics[width=4.8 in]{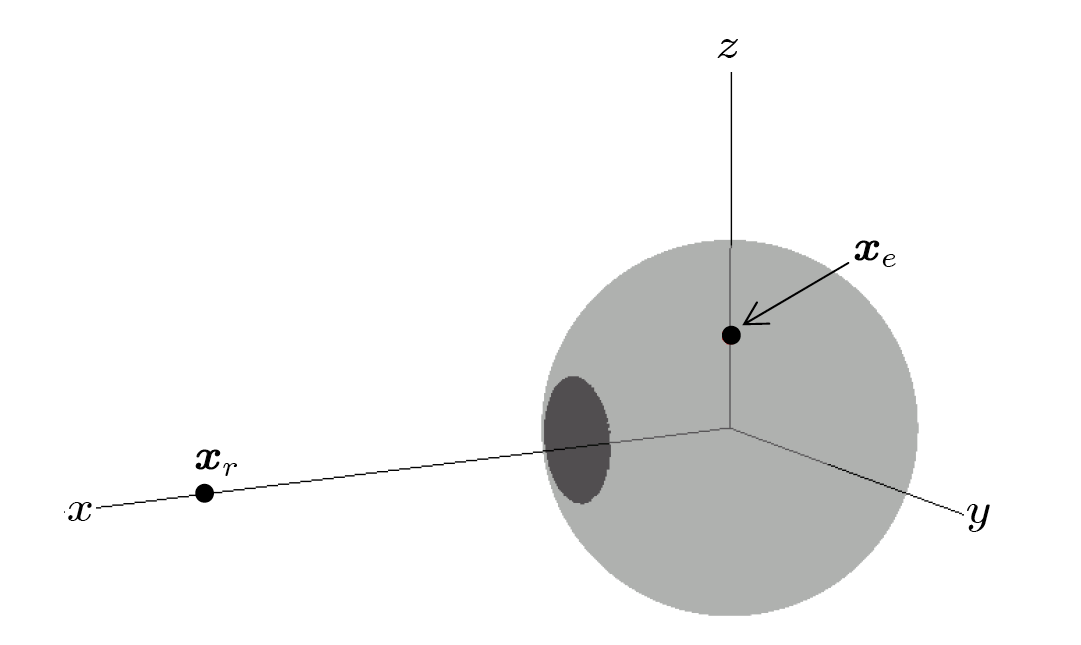}
\caption{\small Decomposition \eq{S2} with
$R=5, a=1,\3x_e=(0, 0,2.5),\3x_r=(20,0,0)$.
The dark region (closest to $\3x_r$) is $S_R^+$ and the light region is $S_R^-$. 
}
\label{F:Fig_eta.png}
\end{center}
\end{figure}

Care must be taken when transforming \eq{rep4} to the time domain because the integrand can grow exponentially in $\o$. Letting
\begin{align}\lab{zxh}
\z=\z_r+\z_e=\x-i\h,\qq \x=\x_r+\x_e,\qq \h=\h_r-\h_e\,,
\end{align}
the exponential in  \eq{rep4} is
\begin{align}\lab{eee} 
e^{i\o\z}=e^{i\o\x}e^{\o\h}.
\end{align}
Setting $\a'=\a$ (as it will be after applying  $\pl_{\a'\a}$), \eq{etaBounds} shows that the bounds on $\h$, as $\bh n$ varies with $\3x_e$ fixed, are
\begin{align}\lab{eta}
-2a\le\h\le (1-\g)a\ \ \hbox{where}\ \  \g=\sr{1-\frac{|\3x_e|^2}{R^2+a^2}}\,.
\end{align}
The upper bound of $\h$ is therefore positive whenever $\3x_e\ne\30$.
This divides the sphere $S_R$ into the subsets 
\begin{align}\lab{S2}
S_R^+(\3x_e, \3x_r)&=\{R\bh n: \h_r>\h_e\}\\
S_R^-(\3x_e, \3x_r)&=\{R\bh n: \h_r\le \h_e\}\nt
\end{align}
shown in Figure \ref{F:Fig_eta.png}.
As indicated, these sets depend on $\3x_e$ and $\3x_r$.  We call $S_R^+$ the \sl frontal zone \rm and $S_R^-$ the \sl rear zone \rm of $S_R$ for the given emission and reception points $\3x_e, \3x_r$. Note that the \sl maximal \rm value $\h_e=a$ is attained when $\3x_e$ is in the direction of $-\bh n$, \ie
\begin{align}\lab{weakest}
\h_e=a\imp\3x_e=-|\3x_e|\bh n,
\end{align} 
which gives the \sl weakest \rm contribution. This is a result of the opposite orientations of the reception and emission disks.

To obtain the time-domain version of \eq{rep4} choose a signal $g\0t,$ multiply both sides by $\1g\0\o$, and take the inverse Fourier transform. Formally, this gives
\begin{align}\lab{rep7}
\frac{g(t-r)}r=
\frac{\a^2}{8\p^2}\,\pl_{\a'\a}\int\frac{\dd\bh n}{\z_r\z_e}\,\ir\dd\o\,e^{-i\o(t-\z)}\1g\0\o,
\end{align}
where we have exchanged the order of integration on the right side, which is justified if the double integral converges absolutely. If $g$ is real, it suffices to compute its positive-frequency component and then take the real part. The positive-frequency component of $g\0t$ is called its \sl analytic signal: \rm
\begin{align}\lab{as}
\2g\0t=\frac1{2\p}\int_0^\8\dd\o\, e^{-i\o t}\1g\0\o,\qq \hat{\tilde g}\0\o=H\0\o\1g\0\o
\end{align}
where $H\0\o$ is the Heaviside step function.
Taking the complex conjugate and using the reality condition $\1g\0\o^*=\1g(-\o)$ gives the negative-frequency component,
\begin{align*}
\2g\0t^*=\frac1{2\p}\int_0^\8\dd\o\, e^{i\o t}\1g(-\o)=\frac1{2\p}\int_{-\8}^0\dd\o\, e^{-i\o t}\1g(\o),
\end{align*}
hence
\begin{align}\lab{re2g}
g\0t=2\re\2g\0t.
\end{align}
If $\1g\0\o$ decays sufficiently rapidly as $\o\to\8$, then the integral
\begin{align}\lab{2gtau}
\2g\0\t=\frac1{2\p}\int_0^\8\dd\o\, e^{-i\o \t}\1g\0\o,\qq \t=t+is
\end{align}
defines an analytic function of the complex time $\t$. The domain of analyticity depends on the decay properties of $\1g$ and the value of $s$. Formally, the positive-frequency part of \eq{rep7} is therefore
\begin{align}\lab{rep7pos}
\frac{\2g(t-r)}r=
\frac{\a^2}{4\p}\,\pl_{\a'\a}\int\frac{\dd\bh n}{\z_r\z_e}\,\2g(t-\z)
\end{align}
provided the integral \eq{2gtau} defining $\2g(t-\z)$ converges absolutely for all $\bh n$. 
Of special interest is the \sl impulse \rm
\begin{align*}
g\0t=\d\0t\imp\1g\0\o\equiv 1\imp\2g\0\t=\frac1{2\p i\t},\qq s<0.
\end{align*}
The integral converges to the \sl Cauchy kernel \rm for $s<0$ and diverges for $s>0$.
The choice $g\0t=\d\0t$ is very attractive since 
\begin{align}\lab{P}
\frac{\d(t-r)}r\equiv P\xt
\end{align}
is the \sl retarded wave propagator, \rm the unique \sl causal fundamental solution \rm of the wave equation:
\begin{align}\lab{fund}
\Box P\0x\equiv (\pl_t^2-\grad^2)P\0x=4\p\d\0x,\qq x=\xt.
\end{align}
$P$ represents the wave radiated by the point source $\d\0x$ at the origin of spacetime. It is `fundamental' because it generates the field radiated by a general source $\vr$ through
\begin{align}\lab{gensol0}
F(x_r)=\int\dd^4 x_e\,P(x_r-x_e)\vr(x_e)\imp\Box F\0x=4\p\vr\0x.
\end{align}
Thus, if we could obtain a pulsed-beam expansion for $P$, this would immediately give a similar expansion for all radiation fields $F$. However, it turns out that the divergence of \eq{2gtau} for $s>0$ makes this task very difficult. Equation \eq{rep7pos} requires
\begin{align*}
\2g(t-\z)=\2g(t-\x+i\h)
\end{align*}
both when $\h_r\le\h_e$ and $\h_r>\h_e$. Numerical experiments have shown that while \eq{rep7pos} `almost' works with the Cauchy kernel, there is always a small but critical \sl failure interval \rm $T=[t_1, t_t]$ where it fails to converge.

Note that disks are ideal for radiating \sl beams \rm (hence we have dish antennas), and
recall that each point on $S_R$ represents a tangent disk of radius $a$. Thus it is reasonable to try constructing a \sl compressed \rm representation \rm of radiation fields by \sl boosting \rm contributions from the frontal zone $S_R^+$, where $\h_r>\h_e$,  and \sl suppressing \rm contributions from the rear zone \rm $S_R^-$, where $\h_r\le\h_e$. The main contributions to \eq{rep7pos} then come from the frontal zone, and this justifies the name `compression.'

However, the Cauchy kernel does this \sl too \rm well: it not only boosts contributions from the frontal zone; it makes them \sl infinite, \rm thus destroying our representation.
We shall solve this problem with an elegant \sl regularization \rm which behaves naturally with respect to spacetime convolutions. This is very important because Huygens' principle is based on spacetime convolutions, as we shall see. Let
\begin{align}\lab{gb}
g_d\0t=\frac{e^{-t^2/d^2}}{\sr{\p}\,d},\qq d>0.
\end{align}
This is the Gaussian distribution with standard deviation $\s=d/\sr{2}$.
Although it seems that generality is lost by specializing to $g_d$, this is actually not the case because
\begin{align}\lab{gb2d}
d\to 0\imp g_d\0t\to \d\0t.
\end{align}
Therefore every continuous signal can be expressed as the limit of a superposition of translated versions of $g_d$:
\begin{align*}
g\0t=\lim_{d\to0}\ir\dd t'\,g(t')g_d(t-t').
\end{align*}
That is, $g_d\0t$ and its translates form a generalized `basis' for signals.
Define the \sl Gaussian wave propagator \rm
\begin{align}\lab{Pb}
P_d\0x=\frac{g_d(t-r)}r,\qq x=(\3x,t),\ r=|\3x|.
\end{align}
By \eq{gb2d}, $P_d$ converges to the retarded wave propagator as $d\to0$:
\begin{align}\lab{Prt}
\lim_{d\to0}P_d\0x=\frac{\d(t-r)}r=P\0x.
\end{align}
Its source is a `Gaussianized' version of $\d\0x$:
\begin{align}\lab{fund2}
\Box P_d\0x=4\p g_d\0t\d(\3x)\equiv 4\p\d_d\0x,\qq\lim_{d\to0}\d_d\0x=\d\0x.
\end{align}
Just as $P$ generates all radiation fields $F$ by \eq{gensol0}, so does $P_d$ generate their Gaussianized versions:
\begin{align}\lab{gensol}
F_d(x_r)\equiv \int\dd^4 x_e\,P_d(x_r-x_e)\vr(x_e),\qq\lim_{d\to0}F_d\0x=F\0x,
\end{align}
whose source is a Gaussianized version $\vr_d$ of $\vr$:
\begin{align*}
\Box F_d\0x&=4\p\int\dd^4 x_e\,\d_d(x_r-x_e)\vr(x_e)\equiv 4\p\vr_d\0x,\qq\lim_{d\to0}\vr_d\0x=\vr\0x.
\end{align*}

\begin{figure}[ht]
\begin{center}
\includegraphics[width=3 in]{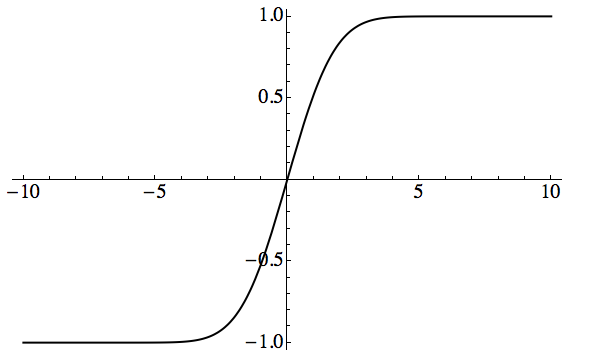}
\caption{\small The error function $\erf(s/d)$, here plotted with $d=2$, is a smoothed version of $\sgn s$ to order $d$. As $d\to 0$, $\erf(s/d)\to\sgn\0s$.}
\label{F:Fig_erf.png}
\end{center}
\end{figure}

The Fourier transform of $g_d\0t$ is
\begin{align*}
\1g_d\0\o=e^{-d^2\o^2/4},
\end{align*}
thus
\begin{align}\lab{gbt}
g_d\0t=\frac1{2\p}\ir\dd\o\,e^{-i\o t}e^{-d^2\o^2/4}.
\end{align}
Both sides extend analytically to the whole complex time plane, giving a Fourier representation of the entire-analytic function $g_d\0\t$:
\begin{align*}
\frac1{2\p}\!\!\ir\dd\o\,e^{-i\o\t}e^{-d^2\o^2/4}&=\frac{e^{-\t^2/d^2}}{2\p}\!\!\ir\dd\o\, e^{-(d\o/2+i\t/d)^2}\\
&=\frac{e^{-\t^2/d^2}}{\sr{\p}\,d}=g_d\0\t.
\end{align*}
The positive-frequency part of $g_d$ is
\begin{align}\lab{2gb0}
\2g_d\0\t=\frac1{2\p}\int_0^\8\dd\o\,e^{-i\o\t}e^{-d^2\o^2/4}
=\frac{e^{-\t^2/d^2}}{2\p}\int_0^\8\dd\o\,e^{-(d\o/2+i\t/d)^2}.
\end{align}
Thus
\begin{align}\lab{2gb}
\2g_d\0\t=\frac12 \, \erfc(i\t/d)g_d\0\t=\frac{w(-\t/d)}{2\sr{\p}d},
\end{align}
where
\begin{align*}
\erfc(i\t/d)=\frac2{\sr{\p}}\int_{i\t/d}^\8\dd u\,e^{-u^2}=1-\erf(i\t/d)
\end{align*}
is the \sl complementary error function \rm and $w$ is the well-known \sl Faddeeva function. \rm  Since both $g_d$ and $\erfc$ are entire, so is $\2g_d$.  Define the function
\begin{align*}
\2H_d(s-it)=\2H_d(-i\t)\equiv \frac12\erfc(i\t/d),
\end{align*}
so that
\begin{align}\lab{2gbtau}
\bx{\2g_d\0\t=\2H_d(-i\t)g_d\0\t.}
\end{align}
As illustrated in Figure \ref{F:Fig_erf.png}, $\erf(s/d)$ is a smoothed version of $\sgn\0s$:
\begin{align*}
\erf(s/d)\equiv \frac2{\sr{\p}}\int_0^{s/d}\dd u\,e^{-u^2}\sim\sgn\0s,\qq 
\lim_{d\to0}\erf(s/d)= \sgn\0s
\end{align*}
and the smoothing is of order $d$, meaning that
\begin{align*}
s<-d\imp \erf(s/d)\app -1\ \ \hbox{and}\ \ s>d\imp \erf(s/d)\app 1.
\end{align*}

\begin{figure}[t]
\begin{center}
\includegraphics[width=2.4 in]{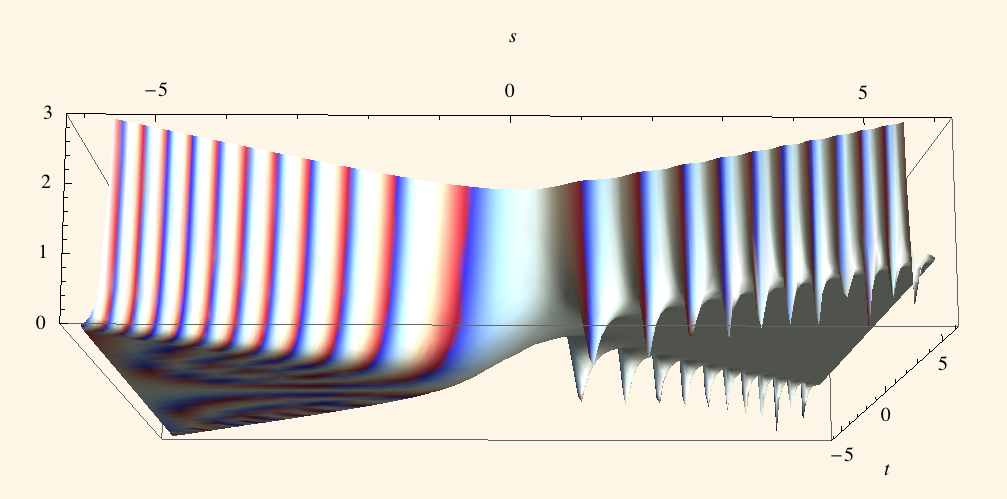}\includegraphics[width=2.4 in]{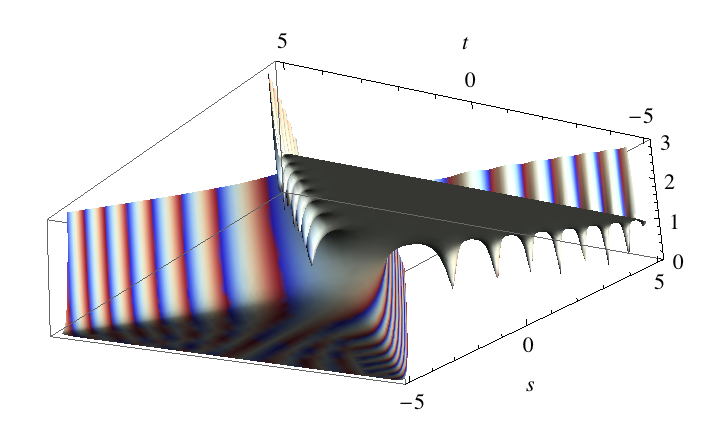}
\caption{\small Plot of $|\2H_d(s-it)|$, shown from the side (left) and from below (right), where it is seen to be an approximation to a smoothed version of $H\0s$ for $|s|>|t|$ and have exponential growth for $|t|>|s|$. The smoothing is of order $d$ and the spikes along $s=|t|$ are zeros. The phase of $\2H_d$ is color-coded on the  surface representing its modus \ci{P9}.}
\label{F:Fig_Htilde.png}
\end{center}
\end{figure}

Since
\begin{align*}
\frac{1+\sgn\0s}2=H\0s=\begin{cases} 1,&s>0\\0,&s<0 \end{cases}
\end{align*}
is the Heaviside step function and
\begin{align*}
2\2H_d(s-it)=1-\erf(i\t/d)=1+\erf((s-it)/d),
\end{align*}
$\2H_d(s-it)$ is the analytic continuation of a smoothed version of $H\0s$ with
\begin{align*}
\lim_{d\to0}\2H_d(s)\to H\0s.
\end{align*}
Again the smoothing is of order $d$:
\begin{align}\lab{Hpm}
s\le -d\imp\2H_d\0s\app 0\ \ \hbox{and}\ \ s\ge d\imp\2H_d\0s\app 1.
\end{align}
For small $d$, $|\2H_d(s-it)|$ is remarkably close to $H\0s$ when $|s|>|t|$. This can be seen in Figure \ref{F:Fig_Htilde.png}. 

\begin{figure}[ht]
\begin{center}
\includegraphics[width=4.8 in]{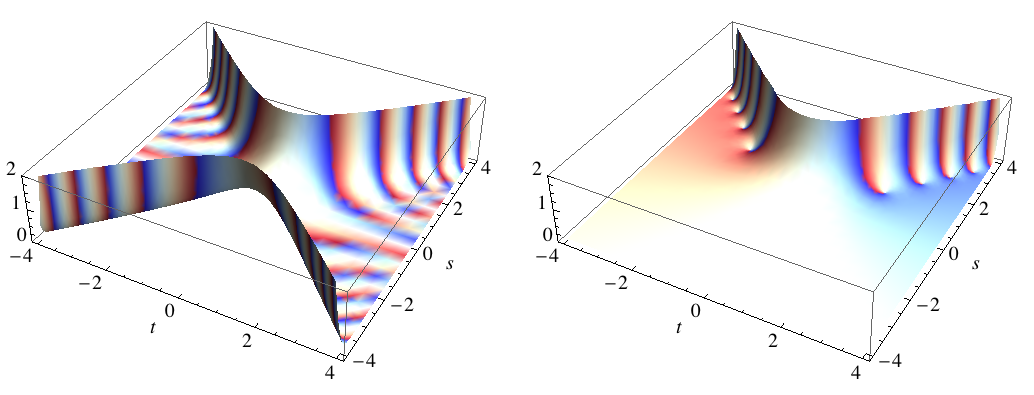}
\caption{\small Plots of $|g_d(t+is)|$ (left) and $|\2g_d(t+is)|$ (right) with $d=1$. $g_d$ grows exponentially in the double cone $|s|>|t|$ and decays exponentially in the double cone $|t|>|s|$ , while $\2g_d$ grows exponentially in the single cone $s>|t|$ and decays elsewhere. The dimples in $\2g_d$ are zeros of $\erfc(i\t/d)$ (see Figure \ref{F:Fig_Htilde.png}) and the phases of $g_d$ and $\2g_d$  are color-coded on the  surfaces representing their moduli \ci{P9}. This shows the oscillation at the compression frequency \eq{compfreq} in the plot of $g_d\0\t$ and its perturbed version (due to the complex factor $\2H_d(-i\t)$) in the plot of $\2g_d\0\t$.}
\label{F:Fig_gbtau.png}
\end{center}
\end{figure}

Furthermore, since
\begin{align*}
\erf(i\t/d)+\erf(-i\t/d)\equiv 0\imp\erfc(i\t/d)+\erfc(-i\t/d)\equiv 2,
\end{align*}
$\2H_d$ extends analytically the \sl partitioning property \rm $H\0s+H(-s)\equiv 1$:
\begin{align}\lab{pu}
\2H_d(-i\t)+\2H_d(i\t)\equiv 1.
\end{align}
Since $g_d\0\t$ is even, \eq{2gbtau} and \eq{pu} imply
\begin{align}\lab{gb3}
\2g_d\0\t+\2g_d(-\t)=g_d\0\t.
\end{align}
But
\begin{align}\lab{gbts}
g_d(t+is)=\frac{e^{(s^2-t^2)/d^2}}{\sr{\p}\,d}\,e^{-2ist/d^2}=g_d\0t e^{s^2/d^2}e^{-2ist/d^2},
\end{align}
hence $g_d(t+is)$ grows exponentially when $|s|>|t|$ and decays exponentially when $|t|>|s|$. The factor $\2H_d(-i\t)$ in \eq{2gbtau} suppresses the negative cone $s<-|t|$, thus making $\2g_d(t+is)$ small everywhere outside the positive cone $s>|t|$.  This is borne out in Figure \ref{F:Fig_gbtau.png}.

More precisely, the continuous-fraction expression \ci[7.1.4]{AS70} for $\erfc$ implies that
\begin{align*}
|\t|\to\8,\ s<0\imp \2g_d(\t)\sim \frac1{2\p i\t},
\end{align*}
hence by \eq{gb3}
\begin{align*}
|\t|\to\8,\ s<0\imp\2g_d(-\t)\sim g_d\0\t- \frac1{2\p i\t}.
\end{align*}
The substitution  $\t\to-\t$  gives 
\begin{align*}
|\t|\to\8,\ s>0\imp \2g_d\0\t\sim g_d\0\t+ \frac1{2\p i\t},
\end{align*}
and the two estimates can be combined into one that will be very useful,\footnote{Equation \eq{asym} is valid for $s=0$ since $g_d\0t\to0$ as $|t|\to\8$.
}
\begin{align}\lab{asym}
\bx{|\t|\to\8 \imp \2g_d\0\t\sim H\0s g_d\0\t+ \frac1{2\p i\t}.}
\end{align}
The `small' value of $\2g_d\0\t$ in the region $s<|t|$ for large $|\t|$ is therefore the Cauchy kernel.
Note that
\begin{align}\lab{2gb1}
\2g_d(-\t)=\frac1{2\p}\int_0^\8\dd\o\,e^{i\o\t}e^{-d^2\o^2/4}
=\frac1{2\p}\int_{-\8}^0\dd\o\,e^{-i\o\t}e^{-d^2\o^2/4}
\end{align}
is the analytic continuation of the negative-frequency part $\2g_d\0t^*$ of $g_d\0t$, as is also clear from \eq{gb3}.

Equation \eq{2gbtau} is remarkable. It shows that $\2H_d(-i\t)$ projects out exactly the positive-frequency part of $g_d\0\t$ by multiplication in the \sl complex time domain, \rm precisely as does $H\0\o$ through \eq{as} in the frequency domain.

\begin{figure}[ht]
\begin{center}
\includegraphics[width=4.5 in]{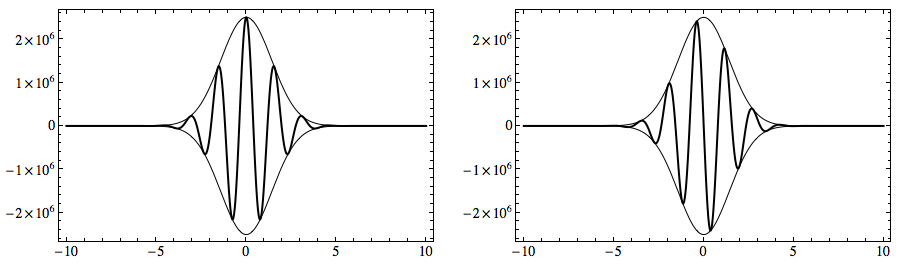}
\caption{\small The real part (left) and imaginary part (right) of $\2g_d(t+is)$ with $d=2$ and $s=8$, together with their envelopes. 
}
\label{F:Fig_ReIm_gb.png}
\end{center}
\end{figure}
Figure \ref{F:Fig_ReIm_gb.png} shows the real and imaginary parts of $\2g_d(t+is)$ as functions of $t$ with given values of $d$ and $s$. They are very similar to those of the real and imaginary parts of $g_d(t+is)$ \eq{gbts}.

With $g=g_d$, the positive-frequency analytic Huygens relation \eq{rep7pos} converges absolutely:
\begin{align}\lab{Huy2g}
\bx{\frac{\2g_d(t-r)}r=\frac{\a^2}{4\p}\,\pl_{\a'\a}\int\frac{\dd\bh n}{\z_r\z_e}\,\2g_d(t-\z).}
\end{align}
By \eq{re2g}, the Gaussian wave propagator \eq{Pb} is given by
\begin{align}\lab{Huyg}
P_d\0x=\frac{g_d(t-r)}r=
2\re\!\!\!\LB\frac{\a^2}{4\p}\,\pl_{\a'\a}\int\frac{\dd\bh n}{\z_r\z_e}\,\2g_d(t-\z)\RB.
\end{align}
Carrying out the differentiations in \eq{Huy2g} gives an expression more suitable for computations:
\begin{align}\lab{Huy3}
\frac{\2g_d(t-r)}r=\frac{\a^2}{4\p}\int\frac{\dd\bh n}{\z_r\z_e}\,
\lb\frac{\z_e'}{\z_e}-\frac{\z_r'}{\z_r}+(\z_e'-\z_r')\pl_t\rb\2g_d(t-\z)
\end{align}
where $\z_r'=\pl_\a\z_r$ and $\z_e'=\pl_\a\z_e$, as in \eq{zrep} after setting $\a'=\a$.
The derivative $\pl_\t\2g_d(t-\z)$ is easily computed. Since
\begin{align*}
\pl_\t\2H_d(-i\t)=\frac1{\sr{\p}}\pl_\t\int_{i\t/d}^\8\dd u\,e^{-u^2}=\frac{e^{\t^2/d^2}}{i\sr{\p} d}
\ \ \ \hbox{and}\ \ \ \pl_\t g_d\0\t=-\frac{2\t}{d^2} g_d\0\t,
\end{align*}
we have
\begin{align}\lab{gdprime}
\pl_\t\2g_d\0\t&=g_d\0\t\pl_\t\2H_d(-i\t)+\2H_d(-i\t)\pl_\t g_d\0\t
=-\frac{2\t}{d^2}\LB\2g_d\0\t-\frac1{2\p i\t}\RB.
\end{align}
Note that
\begin{align*}
|\t|\to\8\ \ \hbox{with}\ \ |t|>s\imp\pl_\t\2g_d\0\t=\5O(\t^{-2})
\end{align*}
because the Cauchy kernel is canceled by \eq{asym} and the next term in the asymptotic expansion of $\2g_d\0\t$ is $\5O(\t^{-3})$.
Inserting this into \eq{Huy3} and using \eq{zrep} gives an expression without any derivatives, ideal for numerical computations.

We shall now interpret \eq{Huyg} as a representation of $P_d$ by a sum of \sl pulsed-beam wavelets \rm radiated by the disks $\5D(\a\bh n)$ tangent to the sphere $S_R$.
It suffices to work with the positive-frequency part \eq{Huy2g}.  Recall that 
\begin{align*}
x=x_r-x_e=(\3x_r-\3x_e, t_r-t_e)=(\3x,t)
\end{align*}
represents the spacetime 4-vector from the \sl emission event \rm $x_e$ to the \sl reception event \rm $x_r$. Consider the \sl intermediate complex event \rm given by\footnote{Recall that $\z_e=\z_e\0\a$ and $\z_r=\z_r(\a')$ and we set $\a'=\a$ after applying $\pl_{\a'\a}=\pl_{\a'}-\pl_\a$.
}
\begin{align}\lab{taup}
z=(\a'\bh n, \t),\ \ \hbox{where}\ \  \t=t_e+\z_e
\end{align}
is the emission time $t_e$ plus the complex travel time $\z_e$ from $\3x_e$ to $\a\bh n$. Define the \sl Gaussian pulsed-beam propagator \rm from $z$ to $x_r$ by
\begin{align}\lab{2Pdz}
\bx{\2P_d(x_r-z)=\frac{\2g_d(t_r-\t-\z_r)}{\z_r}=\frac{\2g_d(t-\z)}{\z_r},\qq
t=t_r-t_e,\ \z=\z_r+\z_e.}
\end{align}
This represents the complex wave amplitude radiated by $\a'\bh n$ at the complex time $\t$ and received at $\3x_r$ at time $t_r$. Thus \eq{Huy2g} reads
\begin{align}\lab{Huyg1}
\2P_d(x_r-x_e)=\frac{\a^2}{4\p}\,\pl_{\a'\a}\int\dd\bh n\,\2P_d(x_r-z)\frac1{\z_e}.
\end{align}
The  general Gaussianized solution $F_d(x_r)$ in \eq{gensol} is therefore given by
\begin{align}
F_d(x_r)&=2\re\2F_d(x_r)\ \ \hbox{where}\ \ \nt\\
\2F_d(x_r)&=\frac{\a^2}{4\p}\,\pl_{\a'\a}\int\dd\bh n\int\dd x_e\,\2P_d(x_r-z)\frac{\vr(x_e)}{\z_e}.
\lab{gensol2}
\end{align}
More will be said about pulsed-beam representations of general solutions in Section \ref{S:GenSol}.

\begin{figure}[ht]
\begin{center}
\includegraphics[width=4 in]{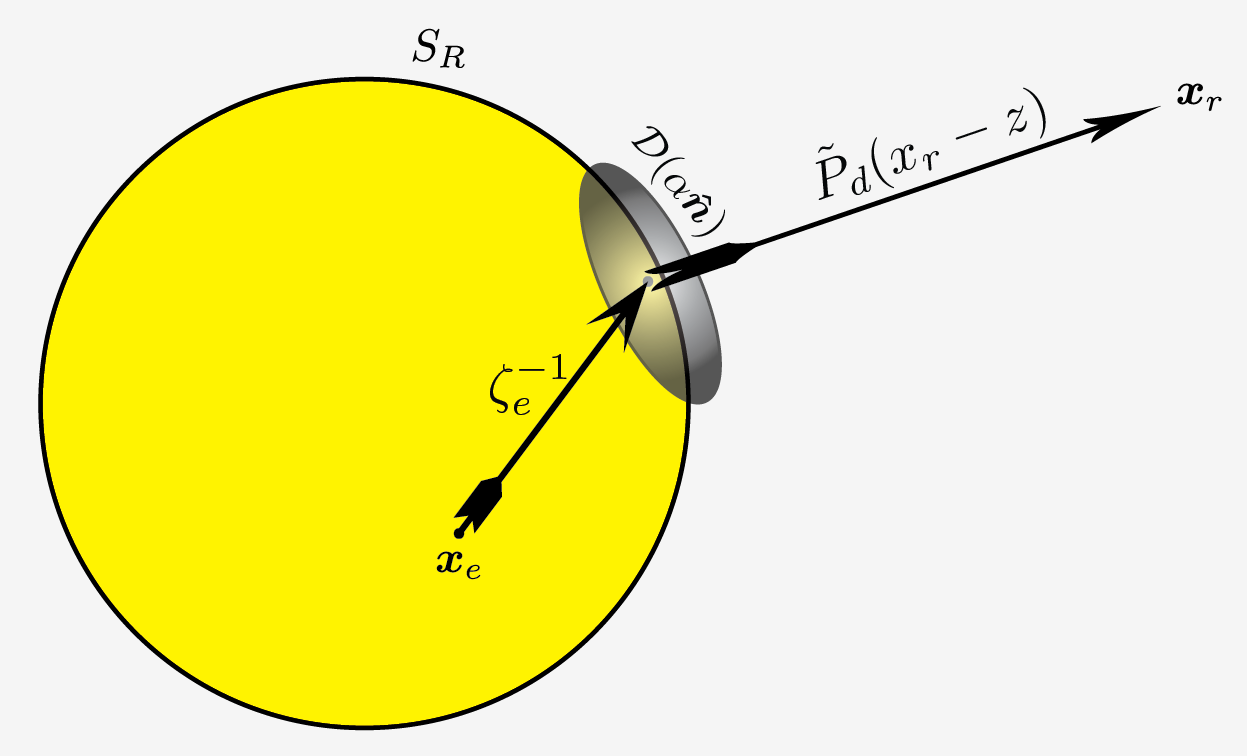}
\caption{\small The factor $\z_e\inv$ in \eq{Huyg1} represents the reception amplitude at $\a\bh n$ due to the attenuation suffered in propagating from $\3x_e$ to $\a\bh n$, and $\2P_d(x_r-z)$ represents the propagation of a pulsed beam from the disk $\5D$ to $\3x_r$.
}
\label{F:Fig_prop.png}
\end{center}
\end{figure}

We now show that $\2P_d(x_r-z)$ is a \sl pulsed beam \rm radiated by $\5D(\a'\bh n)$ which propagates along $\bh n$, with the propagation along $-\bh n$ suppressed by $\2g_d$ as in Figure \ref{F:Fig_gbtau.png}. The factor $\z_e\inv$ represents the  \sl attenuation \rm suffered by the spherical wave emitted by the point source at $\3x_e$ while propagating to $\a\bh n$. Thus we have a picture, shown in  Figure \ref{F:Fig_prop.png}, of a spherical wave emitted at $\3x_e$ and \sl received \rm at $\a\bh n$ with `reception amplitude' $\z_e\inv$, then re-radiated from $\a'\bh n$ as a pulsed beam and finally received at $\3x_r$.\footnote{We are ignoring the derivatives $\pl_{\a'\a}$, so this interpretation is somewhat schematic.
}
The idea of analytically continued fields as reception amplitudes by complex-source disks will be explained in more detail in Section \ref{S:RecAmp}.

By \eq{2Pdz},
\begin{align}\lab{2Pd}
\2P_d(x_r-z)=\frac{\2g_d(t-\z)}{\z_r}=\frac{\2g_d(t-\x+i\h)}{\z_r}.
\end{align}
The properties established for $\2g_d$ show that the magnitude of $\2P_d(x_r-z)$ is an increasing function of $\h$ that attains its maximum values in the frontal zone $S_R^+$ nerest to $\3x_r$; see Figure \ref{F:Fig_eta.png}. Some insight can be gained by noting that
\begin{align*}
g_d(t-\z)=g_d(t-\x)e^{\h^2/d^2}e^{-2i\h(t-\x)/d^2}
\end{align*}
and expanding
\begin{align}\lab{Ptil}
\2P_d(x_r-z)=\z_r\inv\2H_d(\h-i(t-\x))e^{\h^2/d^2}g_d(t-\x)e^{-2i\h(t-\x)/d^2}.
\end{align}
\begin{itemize}
\item At a given time $t$, the factor $g_d(t-\x)$ ensures that $\2P_d(x_r-z)$ is concentrated on a shell of thickness $\sim 2d$ around the surface $\x=t$. $\2P_d$ has significant values only when $t$ is in the range of $\x$, which varies with $\bh n\in S^2$ over a positive interval containing the \sl line of sight \rm time $r=|\3x_r-\3x_e|$, the minimum time required to travel from $\3x_e$ to $\3x_r$ at speed $c=1$. If instead we vary $x_r=(\3x_r, t_r)$ but fix $x_e$ and $\bh n$, this means that the oblate spheroid given by
\begin{align*}
\5O_{\x_r}=\5O_{t-\x_e}
\end{align*}
is a \sl wavefront \rm of $\2P_d(x_r-z)$ expanding with $t=t_r-t_e$. This gives a direct meaning to $\x_r$: \sl it is a variable whose level surfaces are wavefronts. \rm
\item From the behavior of $\2g_d$, it follows that the factor
$\2H_d(\h-i(t-\x)) e^{\h^2/d^2}$
in \eq{Ptil} is an increasing function of $\h$ that boosts \rm the incoming wave when $\h>0$ while suppressing it when $\h<0$.
\item Due to the factor $e^{-2i\h(t-\x)/d^2}$, $\2P_d(x_r-z)$ oscillates at the compression frequency
\begin{align}\lab{compfreq}
\o_d\0\h=\frac{2\h}{d^2}\,,
\end{align}
which depends on $\3x_r$ for given $\bh n$ and on $\bh n$ for given $\3x_r$. This is perturbed slightly by the phase of $\2H_d(\h-i(t-\x))$.
\end{itemize} 

By interpreting every factor in \eq{Ptil}, we have thus understood  $\2P_d(x_r-z)$ as a \sl pulsed beam with wavefronts $\5O_{\x_r}$ propagating along the semi-hyperboloid $\5H_{\h_r}$ at the compression frequency $\o_d$. \rm

Consider the limit of \eq{Huyg1} as $a, a'\to 0$:
\begin{align*}
\2P_d(x_r-x_e)=\frac{R^2}{4\p}\,\pl_{R'R}\int\dd\bh n\,\2P_d(x_r-x)\frac1{r_e},
\end{align*}
where $x=(R'\bh n, t_e+r_e)$ is a reception event on $S_{R'}$ at the arrival time $t_e+r_e$ of a spherical wave radiated from $\3x_e$ at $t_e$. As a function of $x_r$,  $\2P_d(x_r-x)$ is the positive-frequency part of a real Huygens wavelet emitted from $x$.\footnote{Although $P_d(x_r-x)=2\re\2P_d(x_r-x)$ is not a \sl wave \rm because $g_d(t-r)$ does not oscillate, applying the derivative $\pl_{\a'\a}$ gives it some oscillation. For example, 
\begin{align*}
\pl_\a g_d(t-\z)=-\z_e'\pl_t g_d(t-\z)=\frac{2\z_e'}{d^2} (t-\z)g_d(t-\z)
\end{align*}
is a one-cycle wave.
}
By complexifying the sphere, we have deformed the original \sl spherical \rm Huygens wavelets
to pulsed beams \eq{2Pdz}:
\begin{align*}
\2P_d(x_r-x)\to\2P_d(x_r-z).
\end{align*}
This deformation acts on space so that spheres become oblate spheroids and cones become semi-hyperboloids, as in \eq{OH}. In the process of being deformed, the spherical Huygens wavelets are \sl compressed \rm in the forward direction and \sl stretched \rm in the backward direction.\footnote{In a certain sense, they are \sl Doppler scaled \rm positively in the forward direction and negatively in the backward direction \ci{K94}.
}
Being complex, the compression introduces a phase which gives a measure of its strength. This is why we call $\o_d$ the `compression frequency.' Note that
\begin{align*}
a,a'\to0\imp\h\to0\imp\o_d\to0
\end{align*}
as expected. 

\begin{figure}[ht]
\begin{center}
\includegraphics[width=4.5 in]{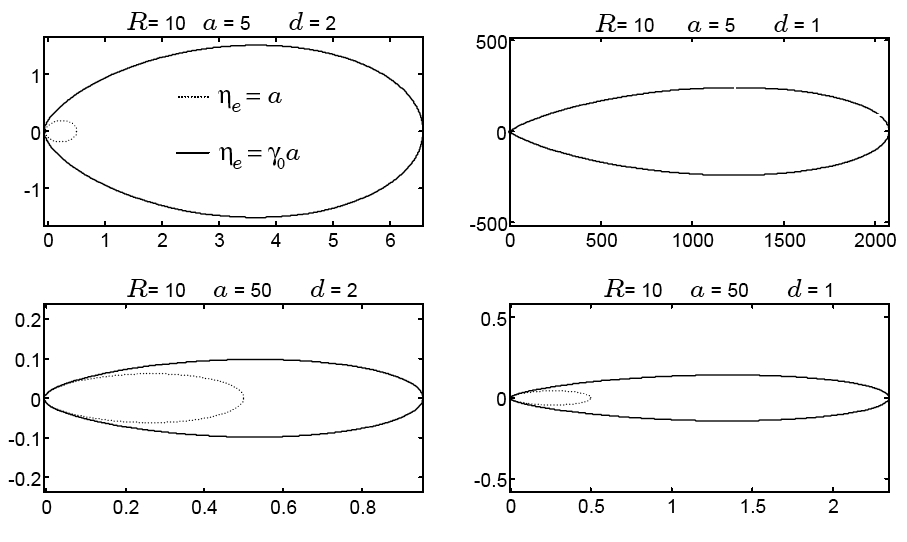}
\caption{\small Peak-time radiation patterns of $\2P_d(x_r-z)$ for $d=1,2$ and $a=5,50$.  In each case we have plotted the beams with the \sl weakest \rm pattern ($\h_e=a$) and the \sl strongest \rm pattern ($\h_e=\g\60 a$). For $a=5$ and $d=1$, the weakest pattern is so weak that it cannot be seen.}
\label{F:Fig_beams0.png} 
\end{center}
\end{figure}

The \sl time-domain radiation pattern \rm of a radiation field $F(\3x,t)$ with cylindrical symmetry is, by definition \ci{HY99},
 the function $\5F(\q, t)$ satisfying the far-field relation
\begin{align*}
F(\3x,t)\sim\frac{\5F(\q, t-r)}r.
\end{align*}
To compute the radiation pattern of $\2P_d(x_r-z)$ \sl relative to the coordinate system of the disk \rm $\5D(\a'\bh n)$, assume the observer is far from the disk.\footnote{Since we are keeping $\bh n$ fixed but varying $\3x_r$, it is unnecessary to assume that $r_r\gg R$. Only the relative vector $\3r_r=\3x_r-R\bh n$ enters the above discussion.
}
Taking $\a'=\a$ for simplicity, \eq{far0} gives
\begin{align*}
r_r\gg a\imp\x_r\sim r_r,\qq \h_r\sim a\bh r_r\cdot\bh n\equiv  a\cos\q_r,
\end{align*}
so that
\begin{align*}
\x\sim r_r+\x_e,\qq \h\sim a\cos\q_r-\h_e.
\end{align*}
The factor $\z_r\inv$ in \eq{2Pdz} can be approximated by $r_r\inv$ since $r_r\gg a$ and $\x_e\le r_e<2a$. Thus
\begin{align*}
\2P_d(x_r-z)\sim\frac{\2g_d(t'-r_r+i(a\cos\q_r-\h_e))}{r_r}\ \ \hbox{where}\ \ t'=t-\x_e.
\end{align*}
Hence the radiation pattern of $\2P_d(x_r-z)$ is
\begin{align}\lab{radpatPd}
\5F(\q_r,t)&=\2g_d(t'+i(a\cos\q_r-\h_e)).
\end{align}
The \sl peak radiation time \rm is $t'=0$, when $t=\x_e$ is the arrival time at $\5D(\a\bh n)$ of the emitted wave. 
Figure \ref{F:Fig_beams0.png} shows polar plots of the peak-time radiation patterns for two values of $a$ and the the two extreme cases with 
\begin{align*}
\h_e=\g\60 a\ \ \hbox{and}\ \ \h_e=a,
\end{align*}
where 
\begin{align*}
\g\60=\lim_{|\3x_e|\to R}\g=\sr{1-\frac{R^2}{R^2+a^2}}=\frac a{\sr{R^2+a^2}}<1
\end{align*}
as in \eq{gamma0}. This lower bound applies to  \sl every source supported in $S_R$. \rm

Since $|\2g_d(t+is)|$  is an increasing function of $s$, the upper bound $\h_e=a$ is 
expected to produce a weaker pattern than the lower bound $\h_e=\g\60 a$, as already discussed beneath \eq{weakest}. This is borne out in Figure \ref{F:Fig_beams0.png}, 
where the pattern with $\h_e=a$  for $a=5$ and $d=1$ is so weak that it is invisible. On the other hand for $a=50$ the disk is so large as to dwarf the sphere. Since $\g\60\app 0.98$ in this case, there is not a great difference between the lower and upper bounds of $\h_e$. This is the reason why both patterns are visible in the lower figures, and why both are much weaker than the patterns with $a=5$ and $\h_e=\g\60 a$.

We have thus established \eq{Huyg1} as a \sl pulsed-beam representation \rm of $\2P_d(x_r-x_e)$. Since the pulsed beams $\2P_d(x_r-z)$ are deformations of Huygens' spherical wavelets, it is reasonable to call them  \sl pulsed-beam Huygens wavelets. \rm By taking the real part and convolving with a general source $\vr$, as in \eq{gensol}, we obtain a pulsed-beam representation of the Gaussianized version $F_d\0x$ of general solution $F\0x$.

\section{Some remarks}

\begin{enumerate}
\item The radiation patterns of extended sources generally have \sl sidelobes, \rm which are interference patterns between parts of the wave arriving from different parts of the source. Sidelobes of beams often stray widely from the intended direction of propagation, causing problems in applications such as communications, remote sensing, and radar  \ci{S98}. However, note that the radiation pattern \eq{radpatPd} is \sl real \rm at the peak time $t'=0$ and it decays monotonically with increasing  $\q_r$, as confirmed by Figure \ref{F:Fig_beams0.png}. It therefore has no sidelobes, and that makes it potentially very useful. If we include the time-dependence around $t'=0$, the radiation pattern acquires a phase factor $e^{-i\o_d t'}$ and $2\re\5F(\q_r,t)$, the radiation pattern of $P_d(x_r-z)$, acquires sidelobes. But these are confined to the narrow  \sl envelope \rm of $2|\5F(\q_r,t)|$ and do not cause the usual problems.

\item To fully justify the name `pulsed-beam propagator,' consider $\2P_d(x_r-z)$ as a function of $x_r$ with $z$ fixed. It is singular on the branch cut $\5D(\a'\bh n)$ of $\z_r(\a')$ and analytic elsewhere, hence
\begin{align*}
\3x_r\notin\5D(\a'\bh n)\imp \Box_r\2P_d(x_r-z)=0
\end{align*}
where $\Box_r$ is the wave operator with respect to $x_r$.
$\2P_d(x_r-z)$ is therefore the wave radiated by the disk $\5D(\a'\bh n)$. The precise source of 
$\2P_d(x_r-z)$ is a generalized function $\2\d_d(x_r-z)$ supported on $\3x_r\in\5D(\a'\bh n)$; see Equation \eq{pfunds} in Section \ref{S:RecAmp}.

\item Taking the complex conjugate of \eq{Huy2g} and substituting $\a,\a'\to\a^*,\a'^*$ (which is permitted since the left side is independent of $\a$ and $\a'$ and the domain \eq{A} is symmetric under conjugation) gives the negative-frequency component in the form
\begin{align}\lab{rep2gbc}
\frac{\2g_d(t-r)^*}r=\frac{\a^2}{4\p}\,\pl_{\a'\a}\int\frac{\dd\bh n}{\z_r\z_e}\,\2g_d(t-\z^*)^*,
\end{align}
where we have used the reality conditions \eq{reality}
\begin{align*}
\z_e(\a^*)^*=\z_e\0\a,\qq \z_r(\a'^*)^*=\z_r(\a').
\end{align*}
Thus $\2g_d\0\t$ satisfies the reality condition
\begin{align*}
\2g_d(\t^*)^*=\2g_d\0\t,
\end{align*}
which also follows directly from \eq{2gb0}. Adding \eq{Huy2g} and \eq{rep2gbc} gives an \sl alternative \rm form of the analytic Huygens relation\footnote{Equation (87) can also be obtained directly from \eq{rep7} with $\1g_d\0\o=e^{-d^2\o^2/4}$.
}
\begin{align}\lab{Huygalt}
\frac{g_d(t-r)}r=
\frac{\a^2}{4\p}\,\pl_{\a'\a}\int\frac{\dd\bh n}{\z_r\z_e}\,g_d(t-\z),
\end{align}
which is simpler than \eq{Huyg} as it does not split up the positive and negative frequencies. 
However, we find that while \eq{Huygalt} is numerically valid, it does \sl not \rm lead to a compressed representation of radiation fields. The problem is the substitutions $\a\to\a^*,\a'\to\a'^*$. For $\a=R+ia$ with $a>0$, 
\begin{align}\lab{alphacc}
\a^*\bh n=R\bh n-ia\bh n.
\end{align}
Hence the disk $\5D(\a^*)$, while still tangent to $R\bh n$, radiates a pulsed beam along $-\bh n$, \ie  to the \sl interior \rm of the sphere. Eventually, this beam leaves the sphere and continues to propagate, weakened, in the direction of $-\bh n$; but this is clearly an inefficient way to represent radiation. Although \eq{Huygalt} is \sl mathematically \rm correct in the sense that the integral on the right converges absolutely to $\2P_d(x_r-x_e)$, this inefficiency shows up in the appearance of very large numbers which spoil the compression and easily overwhelm computational software, thus introducing huge errors; see the discussion at the end  Section \ref{S:Num}.

\item In view of the previous remark, we can say that the positive-frequency part of \eq{rep7} is `good' while its negative-frequency part is `bad.' The situation would be reversed for the \sl interior problem, \rm where a source is given outside of $S_R$ and we seek to represent the field inside $S_R$ as a superposition of pulsed beams. The pulsed-beam analysis and synthesis of interior fields is very similar to that of exterior fields and will be treated elsewhere.

\item The pulsed-beam representation \eq{gensol2} of general radiation fields suggests an important application: given a receiver at $\3x_r$, the most significant contributions are expected to come from disks radiating approximately in the direction of $\3x_r$,  whose centers $R\bh n$ are in the frontal zone $S_R^+$. That is, by using only the `relevant' wavelets propagating toward a given observer, we obtain a \sl compressed \rm representation of $F_d(x_r)$. This is discussed in greater detail in Sections \ref{S:Num} and \ref{S:large}.
\end{enumerate}

\section{Huygens reproducing relation for pulsed beams}\label{S:RepRelPBW}

The time-domain version \eq{Huyg1} of the analytic Huygens principle treats emission and reception asymmetrically: propagation from $x_e$ to $z$ is represented by $\z_e\inv$, whereas propagation from $z$ to $x_r$ is represented by $\2P_d(x_r-z)$. In this section we construct a more complete picture of this process which has a detailed and appealing physical interpretation. For this we shall need to Gaussianize both the emission time $t_e$ and the reception time $t_r$. Thus let $d_e$ and $d_r$ be Gaussian duration parameters for $t_e$ and $t_r$ and let
\begin{align*}
d=\sr{d_e^2+d_r^2},
\end{align*}
which will be the duration parameter for the entire transmission process.
Let
\begin{align}\lab{ztau}
z_\a=(\a\bh n, \t),\qq z_{\a'}=(\a'\bh n, \t),\ \ \hbox{where}\ \  \t=t+is
\end{align}
is a free complex time variable. When $\t=t_e+\z_e$, \ie $t=t_e+\x_e$ and $s=\h_e$, \eq{ztau} reduces to \eq{taup}. 
The propagations from $x_e$ to $z_\a$ and $z_{\a'}$ to $x_r$ are governed by
\begin{align*}
\2P_{d_e}(z_\a-x_e)&=\frac{\2g_{d_e}(\t-t_e-\z_e)}{\z_e}\\
\2P_{d_r}(x_r-z_{\a'})&=\frac{\2g_{d_r}(t_r-\t-\z_r)}{\z_r}
\end{align*}
with
\begin{align*}
\2g_{d_e}(\t-t_e-\z_e)&=\frac1{2\p}\int_0^\8\dd\o\,e^{-i\o(t-t_e-\z_e+is)}e^{-d_e^2\o^2/4}\\
\2g_{d_r}(t_r-\t-\z_r)&=\frac1{2\p}\int_0^\8\dd\o\,e^{-i\o(t_r-t-\z_r-is)}e^{-d_r^2\o^2/4}.
\end{align*}
Applying the Fourier transform in $t'=t-t_e$ to the first equation and in $t''=t_r-t$ to the second equation gives
\begin{align}\lab{1gg}
\ir\dd t'\,e^{i\o t'}\2g_{d_e}(t'-\z_e+is)&=H\0\o\,e^{i\o(\z_e-is)}e^{-d_e^2\o^2/4}\\
\ir\dd t''\,e^{i\o t''}\2g_{d_r}(t''-\z_r-is)&=H\0\o\,e^{i\o(\z_r+is)}e^{-d_r^2\o^2/4}.\nt
\end{align}
Multiplying \eq{rep4} by $H\0\o e^{-d^2\o^2/4}=H\0\o^2 e^{-d^2\o^2/4}$ gives
\begin{align*}
&H\0\o\frac{e^{i\o r}}r\,e^{-d^2\o^2/4}
=\frac{\a^2}{4\p}\pl_{\a'\a}\!\!\int\!\!\dd\bh n\!\lb H\0\o\frac{e^{i\o\z_r}e^{-d_r^2\o^2/4}}{\z_r}\rb\!\!
\lb H\0\o\frac{e^{i\o\z_e}e^{-d_e^2\o^2/4}}{\z_e}\rb\\
&\qqq=\frac{\a^2}{4\p}\pl_{\a'\a}\!\!\int\!\!\dd\bh n\!\lb H\0\o\,\frac{e^{i\o(\z_r+is)}e^{-d_r^2\o^2/4}}{\z_r}\rb\!\!\lb H\0\o\,\frac{e^{i\o(\z_e-is)}e^{-d_e^2\o^2/4}}{\z_e}\rb.
\end{align*}
Taking the inverse Fourier transform and writing the time variable as $t_r-t_e$ gives
\begin{align*}
\frac{\2g_d(t_r-t_e-r)}r=\frac{\a^2}{4\p}\,\pl_{\a'\a}\int\dd\bh n\ir\dd t\,\frac{\2g_{d_r}(t_r-\t-\z_r)}{\z_r}\,
\frac{\2g_{d_e}(\t-t_e-\z_e)}{\z_e}.
\end{align*}
We have thus proved the following result.

\thm The Gaussian pulsed-beam propagator $\2P_d$ satisfies the following complex spacetime Huygens reproducing relation:
\begin{align}\lab{repPBW}
\bx{\2P_d(x_r-x_e)=\frac{\a^2}{4\p}\,\pl_{\a'\a}\int\dd\bh n\ir\dd t\,\2P_{d_r}(x_r-z_{\a'})\2P_{d_e}(z_\a-x_e).}
\end{align}
\rm

\bf Remark 1. \rm
The complex spacetime 4-vector $z_\a=(\a\bh n,t+is)$ represents a \sl pulsed receiving disk \rm with a `Gaussian' reception interval $[t-d_e, t+d_e]$, which we denote by\footnote{The role of $s=\im\t$ will be explained below.
}
\begin{align}\lab{PBr}
\5D_{d_e}(z_\a)\equiv \5D(\a\bh n)\times I_{d_e}\0t\subset \rr4
\ \ \hbox{where}\ \ I_{d_e}\0t=[t-d_e,t+d_e].
\end{align}
Just as $\a\bh n \in\cc3$ represents the extended object $\5D(\a\bh n)$ in space, so does $z\in\cc4$ represent the extended object $\5D_{d_e}(z)$ in spacetime. Similarly, $z_{\a'}$ represents a \sl pulsed emitting disk \rm 
\begin{align}\lab{PBe}
\5D_{d_r}(z_{\a'})\equiv \5D(\a'\bh n)\times I_{d_r}\0t.
\end{align}
The relation between pulsed-beam emitters and receivers will be explained in greater detail in Section \ref{S:RecAmp}. 

\bf Remark 2. \rm
Equation \eq{repPBW} has a simple interpretation which, unlike \eq{Huyg1}, treats emission and reception symmetrically. It states that the spherical wave emitted from the point source $x_e$ is received by $\5D_{d_e}(z_\a)$, then immediately re-emitted by $\5D_{d_r}(z_{\a'})$, and finally received at $x_r$. The direct propagator $\2P_d(x_r-x_e)$ is recovered by integrating over all directions $\bh n$ and intermediate times $t$ and then applying $(\a^2/4\p)\pl_{\a'\a}$. 

\bf Remark 3. \rm
The integral over $t$ can be viewed as a \sl contour integral \rm in $\t$, with the left side independent of $s$ due to analyticity. In fact, $s$ is allowed to depend on $\bh n$ (and even on $t$). For a general emission source $\vr_e(x_e)$ and receiving source $\vr_r(x_r)$, where there are additional integrations over $x_e$ and $x_r$, $s$ may also be allowed to depend on $x_e$ and $x_r$.

\bf Remark 4. \rm
The formal symmetry of \eq{repPBW} with respect to emission and reception is of more than purely academic interest. To formulate pulsed-beam representations for the \sl interior field \rm
given an exterior source $\vr$, we must reverse the roles of $x_e$ and $x_r$ and use the \sl advanced \rm wave propagator:
\begin{align}\lab{advprop}
P\0x\to P'\0x=\frac{\d(t+r)}r.
\end{align}
Then  \eq{repPBW} transforms to a pulsed-beam representation of the interior field but \eq{Huyg1} fails to do so. However, to get an \sl efficient \rm representation, we must also replace $\a,\a'$ by their complex conjugates, as discussed beneath \eq{alphacc} since we now want the pulsed beams to propagate \sl inward. \rm
\sv1  

The properties of $\2g_d$ place some practical constraints on $s$. Let
\begin{align}\lab{tauer}
\t_e&=\t-t_e-\z_e=t-t_e-\x_e+i(s-\h_e)\\
\t_r&=t_r-\t-\z_r=t_r-t-\x_r+i(\h_r-s),\nt
\end{align}
so that
\begin{align}\lab{Per}
\2P_{d_e}(z_\a-x_e)&=\frac{\2g_{d_e}(\t_e)}{\z_e}\ \ \hbox{and}\ \ 
\2P_{d_r}(x_r-z_{\a'})=\frac{\2g_{d_r}(\t_r)}{\z_r}.
\end{align}
The compression frequencies of the interior and exterior pulsed beams, defined as in \eq{compfreq}, are
\begin{align}\lab{cfer}
\o_e=\frac{2(s-\h_e)}{d_e^2},\qq \o_r=\frac{2(\h_r-s)}{d_r^2},
\end{align}
and the propagators in \eq{repPBW} will be very small unless
\begin{align}\lab{ore}
\o_e>0,\qq\o_r>0,
\end{align}
respectively. If \sl both \rm inequalities hold, they imply
\begin{align}\lab{sl0}
\h_e< s<\h_r
\end{align}
which is consistent with $\h=\h_r-\h_e>0$.
For example, 
\begin{align}\lab{sGood}
s=\frac{\h_r+\h_e}2\imp \o_e=\frac\h{d_e^2}\ \ \hbox{and}\ \  \o_r=\frac\h{d_r^2}.
\end{align}
Since $\h_r>\h_e$ for the dominant beams, the condition \eq{ore} is indeed satisfied by \eq{sGood}, showing that \eq{sl0} is sufficient as well as necessary. Although our proof of \eq{repPBW} is theoretically valid for all choices of $s$, the computation can be expected to be \sl inefficient \rm for values of $s$ violating \eq{ore}. For example, if $s$ is large and positive, then  $\2P_{d_e}(z_\a-x_e)$ is very large and $\2P_{d_r}(x_r-z_{\a'})$ is very small. Conversely, choosing $s$ large and negative makes $\2P_{d_e}$ very small and $\2P_{d_r}$ very large. Such choices introduce unnecessary \sl noise \rm into the computation, thus reducing its efficiency and even causing errors when the machine capacity is exceeded, which does in fact occur rapidly due to the exponential growth. Note that choosing $s$ in the `good' interval \eq{sl0} makes it dependent on $\bh n$, which is permissible as explained above.

Numerical calculations confirm that all values of $s$ in the interval $[\h_e, \h_r]$ give stable results and that instabilities build up rapidly when $s$ strays outside this interval.

\section{Analytic Huygens relation for general solutions}\label{S:GenSol}

Let $\vr\0x=\vr\xt$ be a time-dependent source distribution bounded in space, and choose $R$ so that $\vr\xt$ is supported in the open ball $|\3x|<R$ at all times.\footnote{This is always possible if $\vr$ is compactly supported in time as well as space. If $\vr$ is spatially bounded at all times but does not remain in a bounded spatial region (for example, if it drifts at some velocity $\3v\ne\30$), then a generalization of \eq{repPBW} based on a spacetime version of Green's second identity must be used.
}
The radiated field $F$ is given by
\begin{align}\lab{F}
F\0x=\int\dd x_e\,P(x-x_e)\vr(x_e)
\end{align}
where $P$ is the retarded wave propagator \eq{P}. Now Gaussianize the emission time $t_e$ by 
\begin{align*}
\vr_{d_e}(\3x_e, t_e)=\ir \dd t_e'\,g_{d_e}(t_e-t_e')\vr(\3x_e, t_e'),\qq d_e>0.
\end{align*}
Then the wave arriving at $x$ is (by the associativity of convolutions)
\begin{align}\lab{Fs}
F_{d_e}\0x=\int\dd x_e\,P(x-x_e)\vr_{d_e}(x_e)=\int\dd x_e\,P_{d_e}(x-x_e)\vr(x_e)
\end{align}
where $P_{d_e}$ is the Gaussian propagator \eq{Pb}. The spatial integral is over the support of $\vr$ in the interior of the sphere $S_R$. Both sides of \eq{Fs} can be continued analytically to the complex spacetime points
\begin{align*}
x\to z_\a=(\a\bh n, t+is)
\end{align*}
since the integration over $\3x_e$ does not encounter any of the branch cuts of $\z_e=\sr{(\a\bh n-\3x_e)^2}$. The analytic continuation of the positive-frequency part $\2F_{d_e}\0x$ of \eq{Fs} is
\begin{align}\lab{Fs2}
\2F_{d_e}(z_\a)=\int\dd x_e\,\2P_{d_e}(z_\a-x_e)\vr(x_e).
\end{align}
Convolving \eq{repPBW} with $\vr$ gives
\begin{align}\lab{Fs3}
\bx{\2F_d(x_r)=\frac{\a^2}{4\p}\,\pl_{\a'\a}\int\dd\bh n\int\dd t\,
\2P_{d_r}(x_r-z_{\a'})\2F_{d_e}(z_\a)}
\end{align}
where the reception time $t_r$ has also been Gaussianized by convolving with $\2g_{d_r}$, so that the total duration parameter is  $d=\sr{d_e^2+d_r^2}$. 

The analytic Huygens relation \eq{repPBW} for propagators thus implies a pulsed-beam representation for arbitrary solutions with spatially bounded sources. As will be explained in Section \ref{S:RecAmp}, the coefficient $\2F_{d_e}(z_\a)$ in this superposition is the \sl reception amplitude \rm of the interior field $F_{d_e}$ by the pulsed disk $\5D_{d_e}(z_\a)$.

\section{Pulsed-beam reception and emission}\label{S:RecAmp}

The purpose of this section is to justify the interpretation of $\2F_{d_e}(z_\a)$ in \eq{Fs3} as the \sl reception amplitude \rm of the field $F$ by the pulsed disk $\5D_{d_e}(z_\a)$ defined in  \eq{PBr}. By
the wave equation
\begin{align*}
\Box F(x)=4\p\vr(x),
\end{align*}
\eq{Fs2} can be written as a relation between $\2F_{d_e}(z_\a)$ and $F(x_e)$,
\begin{align}\lab{Fs4}
\2F_{d_e}(z_\a)=\frac1{4\p}\int\dd x_e\,\2P_{d_e}(z_\a-x_e)\Box_e F(x_e),
\end{align}
where $\Box_e$ is the wave operator in $x_e$ and 
\begin{align*}
z_\a=(\a\bh n,\t)=x+iy,\qq x=(R\bh n, t),\qq y=(a\bh n, s).
\end{align*}
Integrating by parts twice gives\footnote{$\2P_{d_e}(z_\a-x_e)$ is singular when $\3x_e\in\5D(\a\bh n)$, so the right side of \eq{Fs5} must be treated carefully. The wave operator $\Box_e$ acts on $\2P_{d_e}(z_\a-x_e)$ in a \sl distributional \rm sense, just as it acts on $P_{d_e}(x-x_e)$ to give $4\p\d_{d_e}(x-x_e)$. The resulting distribution $\2\d_{d_e}$ can be computed rigorously using the methods developed in \ci{K0,K3,K4,K5,D8} and will be studied in detail elsewhere. Here we explain the main ideas in an intuitive and informal way.
}
\begin{align}\lab{Fs5}
\2F_{d_e}(z_\a)=\frac1{4\p}\int\dd x_e\,\Box_e\2P_{d_e}(z_\a-x_e) F(x_e).
\end{align}
To make sense of this, note that in \sl real \rm spacetime ($y\to0$) we have \eq{fund2}
\begin{align}\lab{funds}
\Box_e P_{d_e}(x-x_e)=4\p g_{d_e}(t-t_e)\d(\3x-\3x_e)\equiv 4\p\d_{d_e}(x-x_e),
\end{align}
whose positive-frequency part is
\begin{align}\lab{pfunds}
\Box_e \2P_{d_e}(x-x_e)=4\p \2g_{d_e}(t-t_e)\d(\3x-\3x_e)\equiv 4\p\2\d_{d_e}(x-x_e).
\end{align}
We now \sl define \rm the source distribution $\2\d_{d_e}(z_\a-x_e)$ of $\2P_{d_e}(z_\a-x_e)$ by extending this to complex spacetime:
\begin{gather}\lab{2dzt}
4\p\2\d_{d_e}(z_\a-x_e)\equiv \Box_e\2P_{d_e}(z_\a-x_e).
\end{gather}
Since $\2P_{d_e}(z_\a-x_e)$ is analytic whenever $\3x_e\notin\5D(\a\bh n)$, it follows from \eq{funds} that $\Box_e\2P_{d_e}(z_\a-x_e)=0$ at such $\3x_e$. Hence the distribution $\2\d_{d_e}(z_\a-x_e)$ is supported in $\3x_e\in\5D(\a\bh n)$ at all times. It is also localized in a `Gaussian' sense in time around the interval $I_{d_e}\0t$, thus it is effectively localized in the pulsed disk $x_e\in\5D_{d_e}(z_\a)$. 

In other words, the wave operator $\Box_e$ ignores all the \sl analytic \rm behavior of $\2P_{d_e}(z_\a-x_e)$ and nails down its \sl singular \rm behavior, consisting of a discontinuity across the branch cut $\5D(\a\bh n)$ and infinity along the branch circle $\pl\5D(\a\bh n)$. All this is possible only in the distributional sense. Combining \eq{Fs5} and \eq{2dzt} gives
\begin{align}\lab{FsF}
\bx{\2F_{d_e}(z_\a)=\int\dd x_e\,\2\d_{d_e}(z_\a-x_e) F(x_e)}
\end{align}
which is the \sl analytic deformation \rm of $F$. Equation \eq{FsF} shows that $\2F_{d_e}(z_\a)$ is the reception amplitude of $F(x_e)$ by the receiving source $\2\d_{d_e}(z_\a-x_e)$, confirming our claim.

\bf Remark. \rm
The distribution $\2\d_{d_e}$ includes a \sl dipole layer, \rm represented by a first-order differential operator acting on $F(x_e)$  \ci{K0,K3}. Consequently, the right side of \eq{FsF} contains the values of both $F$ and its partial derivatives.

\section{Compressed representations of radiation fields}\label{S:Num}

We now demonstrate that the computational properties of the Huygens representation 
depend strongly on the disk radius $a$ and there can be a significant advantage to 
choosing a complex sphere over a real sphere in numerical calculations. 

Throughout this section, the source is at  $\3x_e=(0,0,2.5)$ and has Gaussian 
time dependence $g_d(t)$ with $d=0.3\sr{2}$. The reception point is in the far zone on the positive $x$-axis. We consider three spheres with $R=10$ and disk radii $a=0, 5,50$. For $a=0$, the sphere is real and the pulsed-beam representation reduces to the classical Huygens representation. 

Let
\begin{align}\lab{HuyInt}
\5I^\a_d(\bh n, t)=2\re\LB\frac{\a^2}{4\p\z_r\z_e}
\lb\frac{\z_e'}{\z_e}-\frac{\z_r'}{\z_r}+(\z_e'-\z_r')\pl_t\rb \2g_d(t-\z)\RB,\ \z=\z_r+\z_e,
\end{align}
where the dependence on $\3x_e$ and $\3x_r$ is implicit and $\pl_t\2g_d(t-\z)$ is given by \eq{gdprime}. Then  \eq{Huy3} reads
\begin{align}\lab{good}
P_d\xt=2\re\2P_d\xt=\int\dd\bh n\,\5I^\a_d(\bh n, t),\qq (\3x,t)=x_r-x_e,\ r=|\3x|.
\end{align}
\begin{figure}[htbp]
\begin{center}
\includegraphics[width=4 in]{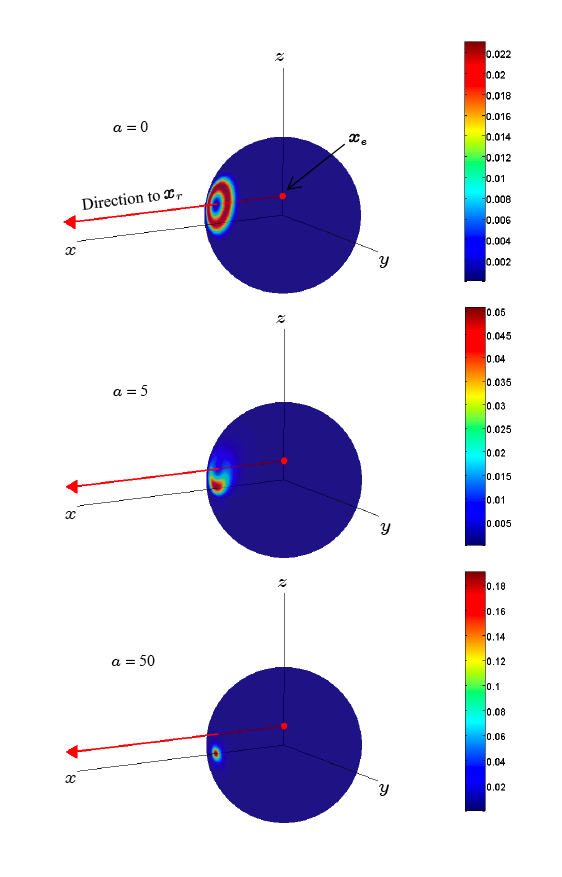}
\caption{\small Magnitudes of $\5I^\a_d$ with $R=10$ at the peak time $t=r$ and $a=0$ (top), $a=5$ (middle), and  $a=50$ (bottom). The emission point is $\3x_e=(0,0,2.5)$ and the reception point is in the far zone ($|\3x_r|\gg|\a|$) on the positive $x$-axis, outside the figure. The line of sight from $\3x_e$ to $\3x_r$ is indicated by the long arrow.}
\label{F:Fig_N1.png}
\end{center}
\end{figure}
\begin{figure}[ht]
\begin{center}
\includegraphics[width=4 in]{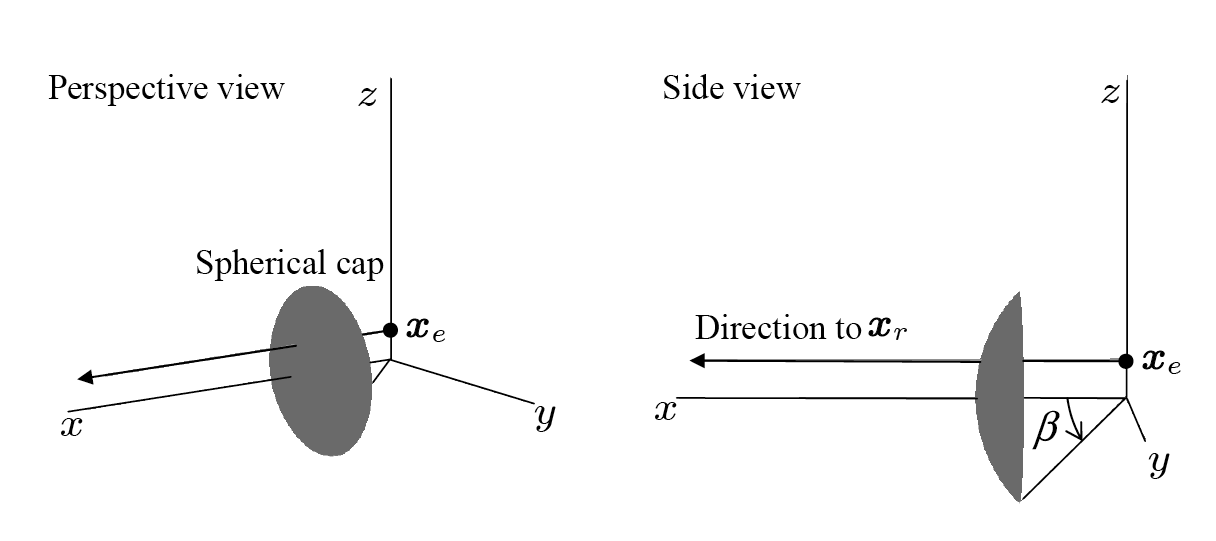}
\caption{\small A spherical cap with cap angle $\b$ centered on the $x$-axis.}
\label{F:Fig_N2.png}
\end{center}
\end{figure}
\begin{figure}[ht]
\begin{center}
\includegraphics[width=4 in]{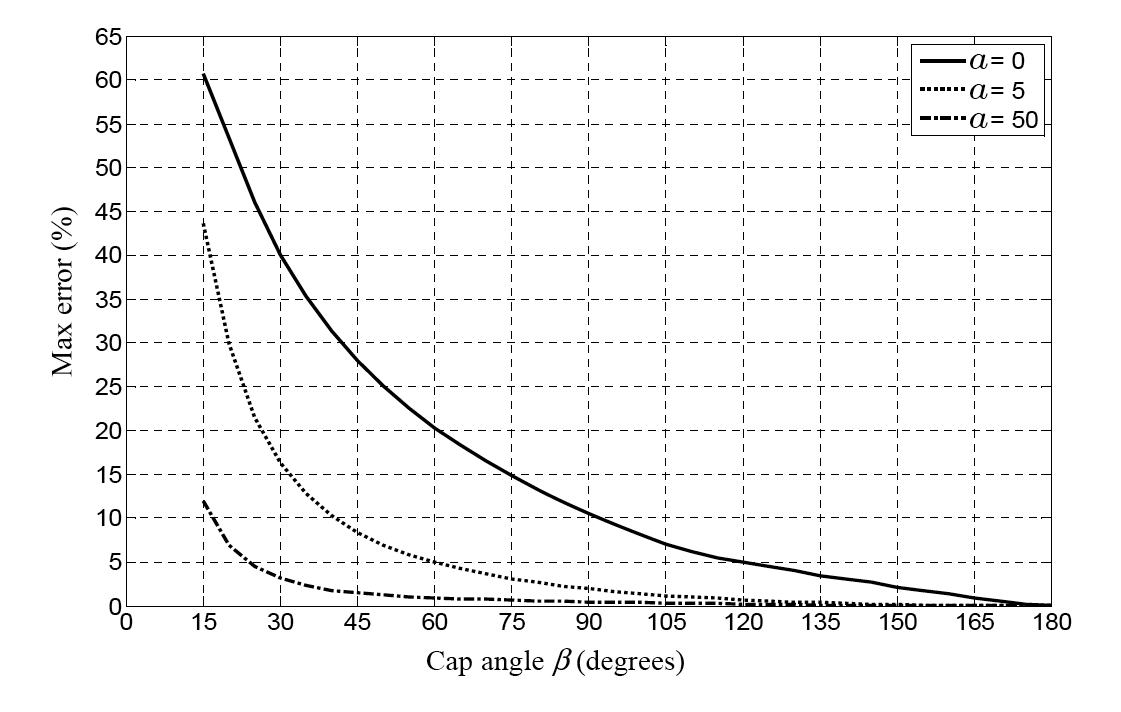}
\caption{\small \bf Compression: \rm The maximum error at a far-zone reception point on the positive $x$-axis as a function of the cap angle $\b$. The emission point is 
on the $\zzz$-axis at $z=2.5$. The three spheres have $R=10$ with disk radii $a=0$,  $a=5$, and $a=50$.}
\label{F:Fig_N3.png}
\end{center}
\end{figure}

Figure \ref{F:Fig_N1.png} shows $\5I^\a_d$ evaluated at the peak time $t=r$, where $P_d\xt$ attains its maximum value $g_d(0)/r$. The top plot shows that $\5I^\a_d$ for the real sphere is significantly nonzero only on a ring centered on the \sl line of sight \rm point from $\3x_e$ to $\3x_r$. We can explain this behavior as follows: for $a=0$,
\begin{align*}
\z_e=|R\bh n-\3x_e|\equiv  r_e\ \ \hbox{and}\ \ \z_r=|\3x_r-R\bh n|\equiv  r_r
\end{align*}
satisfy the triangle inequality
\begin{align*}
r_r+r_e\ge r\equiv |\3x_r-\3x_e|,
\end{align*}
with equality if and only if $R\bh n$ is the line-of-sight point. Now
\begin{align*}
\2g_d(r-\z)=\2H_d(-i(r-r_r-r_e))g_d(r-r_r-r_e)
\end{align*}
decays as a Gaussian in $r-r_r-r_e$ perturbed by $\2H_d(-i(r-r_r-r_e))$. Combined with the effect of the derivatives coming from $\pl_{\a'\a}$, this perturbation displaces the maximum from the line of sight ($r_r+r_e=r$) to a small circle centered at the line-of-sight point. The derivatives cause $\5I^\a_d$ to vanish at the line-of-sight point and oscillate near the circle, thus creating the ring pattern seen in Figure \ref{F:Fig_N1.png}.

For the two complex spheres in the middle and bottom plots, $\2g_d(t-\z)$ suppresses the points $R\bh n$ with $\h_r\le\h_e$ and boosts those with $\h_r>\h_e$ in the frontal zone $S_R^+$ of \eq{S2}. Simultaneously $g_d(r-\x)$ suppresses points with $|r-\x|>d$, and the winners of this tug of war are the points in the small spot centered at $R\bh x_r$ in the middle and bottom plots of Figure \ref{F:Fig_N1.png}. This spot becomes more and more concentrated near $R\bh x_r$ with increasing $a$, as shown in Section \ref{S:large}. 

It is important to note that $\5I^\a_d$ for large $a$ is concentrated around the point $R\bh x_r$ nearest to the receiver  \sl regardless of the location of the emission point. \rm This means that $\5I^\a_d$ for large $a$ would remain concentrated near $R\bh x_r$, as in Figure \ref{F:Fig_N1.png}, even if we replaced the point source at $\3x_r$ with an arbitrary volume source $\vr(x_e)$ supported throughout the interior of $S_R$. This can be understood by noting that a disk source becomes more \sl directive \rm with increasing radius, hence fewer disks are required to achieve a given accuracy.

To illustrate the advantages of using complex spheres with large value of $a$, we investigate the accuracy of the field at $\3x_r$ obtained by including only the pulsed beams radiated from a spherical cap centered around $R\bh x_r$ with maximum angle $\b$, as in Figure \ref{F:Fig_N2.png}.
For the parameter values under consideration, $\b_0=15^\circ$ represents a cap where the line of sight from $\3x_e$ to $\3x_r$ just grazes the upper edge of the cap.  We compute the maximum error over all time of the field at $\3x_r$, relative to the maximum value $g_d(0)/r$. This calculation is designed to simulate the realistic situation where the time dependence of the source is unknown and the source may emit a series of pulses that are spread out over time. In particular, the computed error bounds also hold for time-harmonic fields. Figure \ref{F:Fig_N3.png}  shows this maximum error as a function of $\b\ge \b_0$ for the three values of $a$. A dramatic error reduction is obtained by increasing $a$. For example, the errors $\e(a)$ for $\b=45^\circ$ are
\begin{align*}
\e\00=27.9\%,\qq \e\05=8.3\%,\qq \e(50)=1.5\%,
\end{align*}
a reduction of nearly 20:1 when using the pulsed-beam representation with $a=50$ compared to the real Huygens representation! This has important practical implications in numerical calculations, where a certain error level must be often achieved. For example, assume that the error of the field at the reception point must be less than $2\%$.  Then the required cap angles $\b\0a$ are
\begin{align*}
\b\00=152^\circ,\qq \b\05=89^\circ, \qq\b(50)=38^\circ.
\end{align*}
Hence, with $a=50$ we need to include pulsed-beam wavelets over just $11\%$ of the sphere, whereas $a=0$ would require spherical wavelets over $94\%$ of the sphere. 

\bf Remark. \rm 
Consider the efficiency of the `alternate' expression using \eq{Huygalt}, whose negative-frequency component was obtained by letting $\a\to\a^*$ and $\a'\to\a'^*$ in the conjugate pulsed-beam expansion  \eq{Huy3} of $\2P_d\xt^*$. This gives the expression
\begin{align}\lab{bad}
P_d\xt=\frac{\a^2}{4\p}\int\frac{\dd\bh n}{\z_r\z_e}\,
\lb\frac{\z_e'}{\z_e}-\frac{\z_r'}{\z_r}+(\z_e'-\z_r')\pl_t\rb g_d(t-\z)
\equiv \int\dd\bh n\,\5J^\a_d(\bh n, t),
\end{align}
where $\2g_d$ has been replaced by $g_d$ in the integrand. As noted under \eq{Huygalt}, this means that the negative-frequency pulsed beams propagate along $-\bh n$ instead of $\bh n$, thus traversing the sphere and spoiling the efficiency of \eq{Huyg}. To demonstrate the enormous difference between \eq{good} and \eq{bad}, consider again the complex sphere with $\a=10+50 i$,  $\3x_e=(0,0,2.5)$, and $\3x_r=(200,0,0)$ in the far zone. Since \eq{bad} does not have a factor like $\2H_d$, there will be significant contributions from the back side of the sphere. In particular, consider 
\begin{align*}
\bh n=-\bh x_r=(-1,0,0)\imp \z=\z_r+\z_e=220.01+99.94 i.
\end{align*}
Choosing $d=.3\sr{2}$ as before and $t=\x=220.01$, we find
\begin{align*}
g_d(t-\z)=g_d(-99.94 i)=1.33e^{55489}\sim10^{25000}.
\end{align*}
The contribution from the back point $-\bh x_r$ in \eq{bad} is seen astronomical!
By comparison, the counterpart of $g_d(-99.94 i)$ in \eq{good} is
\begin{align*}
\2g_d(-99.94i)=0.00159.
\end{align*}
This illustrates that \eq{good} and \eq{bad} are on opposite sides of the computational efficiency spectrum.

\section{Pulsed-beam representations for large $a$}\label{S:large}

Figure \ref{F:Fig_N3.png} suggests that the compression keeps improving as $a$ increases, and it is natural to wonder if there is an `optimal' value of $a$ beyond which the benefits of further increase will diminish. We therefore analyze the asymptotics of the pulsed-beam representation for large $a$.

Define $\q,\,\q_e,$ and $\y$ by
\begin{align}\lab{defs}
\cos\q=\bh x_r\cdot\bh n,\qq\cos\q_e=\bh r_e\cdot\bh n,\qq \cos\y=\bh x_r\cdot\bh x_e.
\end{align}
Suppose that $a\gg R$ and that $\3x_r$ is in the far zone \eq{far}:
\begin{align}\lab{large-a}
R\ll a\ \ \hbox{and}\ \ |\3x_r|\gg |\a|.
\end{align}
Since $|\3x_e|<R\ll|\3x_r|$, we have
\begin{align}\lab{r}
r=|\3x_r-\3x_e|=|\3x_r|-\bh x_r\cdot\3x_e+\5O(R/|\3x_r|)\sim |\3x_r|-|\3x_e|\cos\y.
\end{align}
Also
\begin{align*}
\z_e&=\sr{(\3r_e+ia\bh n)^2}
=r_e\cos\q_e+ia\lb 1-\frac{r_e^2\sin^2\q_e}{2a^2}\rb+\5O(R^2/a^2)
\end{align*}
and
\begin{align*}
\z_r&=\sr{(\3x_r-\a\bh n)^2}=|\3x_r|-R\cos\q-ia\cos\q+\5O(\a^2/|\3x_r|),
\end{align*}
thus
\begin{align}\lab{zeta-a} \z\sim r_e\cos\q_e+|\3x_r|-R\cos\q
+ia\lb 2\sin^2(\q/2)-\frac{r_e^2\sin^2\q_e}{2a^2}\rb.
\end{align}
This shows that for large $a$, the frontal zone $\h>0$ is given by 
\begin{align}\lab{frontzone}
\bx{S_R^+\sim\LB R\bh n:\,\sin(\q/2)<\frac{r_e\sin\q_e}{2a}\RB.}
\end{align}
As $a$ increases, $S_R^+$ shrinks to the point $\{R\bh x_r\}$ nearest to the receiver.
At the center of $S_R^+$ we have $\bh n=\bh x_r$ and 
\begin{align*} &r_e^2=(R\bh x_r-\3x_e)^2=R^2+|\3x_e|^2-2R|\3x_e|\cos\y\\
&r_e\cos\q_e=(R\bh x_r-\3x_e)\cdot\bh x_r=R-|\3x_e|\cos\y, 
\end{align*}
hence
\begin{align*} 
\bh n=\bh x_r\imp r_e^2\sin^2\q_e=|\3x_e|^2\sin^2\y\equiv b^2,
\end{align*}
where $b$ is the length of the projection of $\3x_e$ to the plane orthogonal to $\3x_r$.
Therefore
\begin{align*}
\bh n=\bh x_r\imp\h\sim \frac{b^2}{2a}.
\end{align*}
The real part of \eq{zeta-a} gives
\begin{align*}
\bh n\to\bh x_r\imp \x\to (R\bh x_r-\3x_e)\cdot\bh x_r+|\3x_r|-R=| \3x_r|-|\3x_e|\cos\y, \end{align*} thus by \eq{r}, 
\begin{align}\lab{rr} 
\bh n\to\bh x_r\imp \x\to r.
\end{align}

We have thus established that $\zeta=\x-i\h$ behaves as follows as $a$ increases:
\begin{align*}
\bh n\ne\bh x_r&\imp \xi\to r_e\cos\q_e+|\3x_r|-R\cos\q\ \ \hbox{and}\ \ \eta\to-\8\\
\bh n=\bh x_r&\imp \xi\to r\ \ \hbox{and}\ \ \eta\sim b^2/2a.
\end{align*}
The asymptotic formula \eq{asym}  shows that as $a$ increases, $\2g_d(t-\z)\to 0$ outside a shrinking cap centered at $R\bh x_r$ 
while in the center we have 
\begin{align*} 
\bh n=\bh x_r\imp\2g_d(t-\z)\sim \2g_d(t-r+ib^2/2a).
\end{align*}
Figure \ref{F:Fig_gbtau.png} shows that the \sl phase \rm of $\2g_d(t+is)$ remains approximately constant for $s\le 0$. Hence $\2g_d(t-\z)$ decays smoothly without oscillations as  $\bh n$ moves away from $\bh x_r$. Furthermore, it remains bounded throughout the sphere as $a$ grows, and its maximum magnitude approaches $|\2g_d(t-r)|$.  \rm

Thus the compression keeps improving as $a$ increases, and there is no limit to how small the cap angle $\b$ can be for a given error limit. But as $a$ increases, the sampling rate used in the computation of the integral \eq{good} must also be increased to capture the rapid variation of $\2g_d(t-\z)$ near $\bh n=\bh x_r$. We shall investigate the numerical consequences of
using large values of $a$ more fully in future work.

\section*{Acknowledgements}
It is our pleasure to thank Dr.~Arje Nachman and the Air Force Office of Scientific Research for their support of this work.  GK was supported by AFOSR Grant \#FA9550-08-1-0144. We also thank David Park for helping with the Mathematica graphics.

\end{document}